\documentclass{aa}
\usepackage{color}
\usepackage{graphicx}
\usepackage{float}
\usepackage{multirow}
\usepackage{ amssymb }
\newcommand{\hk}{HK\,Tau\,B}
\newcommand{\hv}{HV\,Tau\,C}
\newcommand{\tauZero}{Tau\,042021}
\newcommand{\IRAS}{IRAS\,04302}
\newcommand{\irasGerrit}{IRAS\,04158}
\newcommand{\JDeuxDeux}{Haro~6-5B}
\newcommand{\JTroisZeroSept}{IRAS\,04200}
\newcommand{\hh}{HH\,30}
\newcommand{\Oph}{Oph\,163131}
\newcommand{\Junun}{ESO-H$\alpha$\,569}
\newcommand{\Junsois}{ESO-H$\alpha$\,574}
\newcommand{\hhq}{HH\,48\,NE}

\begin{document} 
   \title{Observations of edge-on protoplanetary disks with ALMA}
    \subtitle{I. Results from continuum data}
   \author{M.~Villenave\inst{1,}\inst{2}  
            \and 
          F.~M\'enard\inst{1}    
          \and 
          W.~R.~F.~Dent\inst{3}
          \and
          G.~Duch\^{e}ne\inst{4,}\inst{1}
          \and
          K.~R.~Stapelfeldt \inst{5}
          \and
          M.~Benisty\inst{6,}\inst{1}
          \and
          Y.~Boehler\inst{1}
          \and
          G.~van~der~Plas\inst{1}
          \and
          C.~Pinte\inst{7,}\inst{1}
          \and 
          Z.~Telkamp\inst{4}
          \and 
          S.~Wolff\inst{8}
          \and 
          C.~Flores\inst{9}
          \and
          G.~Lesur\inst{1}
          \and
          F.~Louvet\inst{10}
          \and
          A.~Riols\inst{1}
          \and
          C.~Dougados\inst{1}
          \and 
          H.~Williams\inst{11,}\inst{4}
          \and
          D. Padgett\inst{5}
    }

   \institute{Univ. Grenoble Alpes, CNRS, IPAG, 38000 Grenoble, France. \\
   			\email{marion.villenave@univ-grenoble-alpes.fr}
             \and 
		      European Southern Observatory, Alonso de C\'ordova 3107, Vitacura, Casilla 19001, Santiago 19, Chile 
            \and
    		Joint ALMA Observatory, Alonso de C\'ordova 3107, Vitacura 763-0355, Santiago, Chile
            \and
            Astronomy Department, University of California, Berkeley, CA 94720, USA
            \and
            Jet Propulsion Laboratory, California Institute of Technology, 4800 Oak Grove Drive, Pasadena, CA 91109, USA
            \and 
            Unidad Mixta Internacional Franco-Chilena de Astronom\'{i}a (CNRS, UMI 3386), Departamento de Astronom\'{i}a, Universidad de Chile, Camino El Observatorio 1515, Las Condes, Santiago, Chile
            \and
            Monash Centre for Astrophysics (MoCA) and School of Physics and Astronomy, Monash University, Clayton Vic 3800, Australia
            \and
            Leiden Observatory, Leiden University, 2300 RA Leiden, The Netherlands
            \and
            Institute for Astronomy, University of Hawaii, 640 N. Aohoku Place, Hilo, HI 96720, USA 
            \and
            AIM Paris-Saclay D\'epartement d'Astrophysique, CEA, CNRS, Univ. Paris Diderot, CEA-Saclay, F-91191 Gif-sur-Yvette Cedex, France
             \and
            School of Physics \& Astronomy, University of Minnesota, 116 Church Street SE, Minneapolis, MN 55455 USA  
		}

   \date{\today}

 \abstract
 {}
 {We aim to study vertical settling and radial drift of dust in protoplanetary disks from a different perspective: an edge-on view. An estimation of the amplitude of settling and drift is highly relevant to understanding planet formation. }
 {We analyze a sample of 12 HST-selected edge-on protoplanetary disks (i.e., seen with high inclinations) for which the vertical extent of the emission layers can be constrained directly. We present ALMA high angular resolution continuum images~($\sim$0.1\arcsec) of these disks at two wavelengths, 0.89\,mm and 2.06\,mm (respectively ALMA bands~7 and~4), supplemented with archival band~6 data (1.33\,mm) where available.}
 {Several sources show constant brightness profiles along their major axis with a steep drop at the outer edges. Two disks have central holes with additional compact continuum emission at the location of the central star.  For most sources, the millimeter continuum emission is more compact than the scattered light, both in the vertical and radial directions. Six sources are resolved along their minor axis in at least one millimeter band, providing direct information on the vertical distribution of the millimeter grains. For the second largest disk of the sample, \tauZero, the significant difference in vertical extent between band~7 and band~4 suggests efficient size-selective vertical settling of large grains. Furthermore, the only Class~I object in our sample shows evidence of flaring in the millimeter. Along the major axis, all disks are well resolved. Four of them are larger in band~7 than in band~4 in the radial direction, and three have a similar radial extent in all bands. These three disks are also the ones with the sharpest apparent edges (between 80\% and 20\% of the peak flux, $\Delta r/r \sim 0.3$), and two of them are binaries. For all disks, we also derive the millimeter brightness temperature and spectral index maps. We find that all edge-on disks in our sample are likely optically thick and that the dust emission reveals low brightness temperatures in most cases (brightness temperatures $\leq$\,10\,K). The integrated spectral indices are similar to those of disks at lower inclination.}
 {The comparison of a generic radiative transfer disk model with our data shows that at least three disks are consistent with a small millimeter dust scale height, of a few\,au~(measured at r=100\,au). This is in contrast with the more classical value of h$_\mathrm{g}\sim$10\,au derived from scattered light images and from gas line measurements. These results confirm, by direct observations, that large (millimeter) grains are subject to significant vertical settling in protoplanetary disks. 
 }

   \maketitle
%-------------------------------------------------------------------

\section{Introduction}
The process of planet formation requires small submicron-sized particles to grow up to large pebbles and boulders that will eventually aggregate to form planetesimals and planets. Given the short lifetimes of disks, such efficient growth has to occur in the highest density regions of protoplanetary disks, that is to say, the inner regions and/or the disk midplane. 
Gas drag, the interaction of dust~(in Keplerian rotation) with the gas orbiting around the central star at slower~(sub-Keplerian) velocities, is at the origin of the vertical settling to the midplane and of the inward radial drift of large~(e.g., millimeter-sized) dust grains~\citep{Weidenschilling_1977, Barriere-Fouchet_2005}.  Unlike the larger grains, micron-sized particles are well coupled to the gas and are located in similar regions, well-mixed with the gas. 
The characteristic timescale of radial drift is predicted to be about a hundred times longer than that of vertical settling~\citep{Laibe_2014c}. However, the strength of these effects is not yet well constrained and depends in particular on the disk viscosity and/or turbulence, on the gas-to-dust ratio, and on the initial grain size distribution~\citep{Fromang_2006, Dullemond_Dominik_2004, Mulders_Dominik_2012}.
The comparison of observations at widely different wavelengths, for example optical-NIR~(near-infrared) and~(sub)millimeter, can help to constrain the settling intensity and radial drift of dust grains. 
Moreover, by performing multi-wavelength observations in the millimeter, one can achieve spectral index measurements and, assuming optically thin emission, interpret the findings in terms of grain growth~\citep[e.g.,][]{Guilloteau_2011, perez_2012}. 

Up to now, most studies of protoplanetary disks have focused on radial features, such as rings, gaps, and spirals, which naturally led to observations of low inclination systems~($\leq70^\circ$), where they are more readily visible. 
Some of these studies showed that the gas is often more radially extended than the millimeter-sized dust \citep[][]{Ansdell_2018, Facchini_2019}, which is likely a combined effect of optical depth and dust radial drift~\citep{Facchini_2017}. However, though it is important to constrain radial drift, dedicated comparisons of the radial distribution of different dust grains sizes remain sparse~\citep[see e.g.,][]{Pinilla_2015, Long_2019}.

For relatively face-on disks, it is difficult to estimate the difference in vertical extent between gas and dust grains. Detailed modeling of ring and gap features is required~\citep[e.g.,][]{Pinte_2016}. 
On the other hand, edge-on disks offer a unique perspective, as they allow us to directly observe their vertical structure. 
Previous studies of edge-on disks at different wavelengths indicate that the grains are stratified, with larger dust closer to the disk midplane~\citep[][]{ Glauser_2008, Duchene_2003, Duchene_2010, Villenave_2019}, as predicted by models. 
However, even for these very inclined systems, the vertical extent of the midplane remains poorly constrained in early studies, limited by the resolution of the observations. 

In this work, we present a survey of edge-on disks observed with the Atacama Large Millimeter Array~(ALMA) at high angular resolution~($\sim$0.1\arcsec). The sample was selected based on Hubble Space Telescope~(HST) images and most of the targets were observed at multiple millimeter bands. After describing the sample and the data reduction in Section~\ref{sec:reduction}, we present the fluxes, brightness temperatures, spectral indices, and the radial and vertical extents of all disks in Section~\ref{sec:results}.  Section~\ref{sec:discussion} compares our results with a toy model. We focus on the vertical and radial extent of the disks, and investigate optical depth in the disks by studying the brightness temperature and spectral indices of the sources. Finally, we summarize our conclusions in Section~\ref{sec:conclusion}.

\section{Observations and data reduction}
\label{sec:reduction}

\subsection{Target selection}
\label{sec:targets}
In this study, we selected a sample of twelve highly inclined disks. All sources were identified as candidates from their Spectral Energy Distribution~\citep[SED, see][and Stapelfeldt et al., in prep]{Stapelfeldt_2014} and confirmed as edge-on disks (hereafter EOD) according to their optical or near-infrared HST scattered-light images.
At these wavelengths, edge-on disks are inclined enough so that the star is not in direct view of the observer.
The ALMA observations targeted 8~sources located in the Taurus star-forming region, 3 in Chamaeleon~I, and 1 in Ophiuchus. 
Most of the observations presented in this work were part of our cycle~4 and~5 survey of edge-on disks (Project 2016.1.00460.S, PI: M\'enard), but we also included complementary observations of \tauZero, \hh, and \Oph\:from previous programs (Projects 2013.1.01175.S, 2016.1.01505.S, and 2016.1.00771.S, PIs: Dougados, Louvet, and Duch\^ene). 

We report the coordinates, spectral types, and stellar masses of the sources in Table~\ref{tab:stellar_parameters}. As the spectral features associated with the central source are still visible for edge-on disks through the scattered light~\citep{Appenzeller_2005}, the spectral types could be determined from spectroscopy~\citep[][Flores et al., in prep]{Luhman_2007, Luhman_2010}. However, the edge-on configuration does not allow a direct estimate of the stellar luminosity. Thus, we report dynamical stellar masses estimated from CO emission. 

At optical and NIR wavelengths, edge-on disk systems highlight extended nebulosities on both sides of a dark lane, tracing the disk midplane. Because of severe extinction in the dark lane, the central star is usually undetected at optical-NIR wavelengths, also resulting in fainter systems for a given spectral type and distance. In a few cases, the brightness distribution of the nebulosities has also been observed to vary significantly~\citep[][]{Stapelfeldt_1999}. These facts combine to render parallax measurements difficult and the distances determined by {\sc {gaia}} can be plagued by large errors. For our targets, we decided to adopt the average distance of the parent star-forming regions instead; 140\,pc for the sources in Taurus and Ophiuchus~\citep{Kenyon_2008, Ortiz-Leon_2018, Canovas_2019} and 192\,pc for those in Chamaeleon~I~\citep{Dzib_2018}. 

Four of the observed disks are part of multiple systems: \hk, \hv, \JDeuxDeux, and~\hhq~\citep{Stapelfeldt_1998, Stapelfeldt_2003, Krist_1998, Haisch_2004} with apparent companion separations larger than 2.4\arcsec. 
While \hh~has been suggested to be a binary on the basis of jet wiggles and a disk central hole in lower resolution and signal-to-noise ratio millimeter continuum maps \citep{Guilloteau_2008}, higher resolution ALMA observations do not confirm the central hole \citep[][]{Louvet_2018}. Neither do they exclude the possibility of spectroscopic binary. The other targets in the sample are not known to be in multiple systems.

\begin{table*}[]
    \centering
    \caption{Stellar parameters.}
    \begin{tabular}{llccccc}
        \hline
        \hline \noalign{\smallskip}
        Full name& Adopted name &RA (h m s)& Dec ($^\circ$ \arcmin~ \arcsec)& SFR & SpT$^{a,b}$ & M$_\star^{~c}$ (M$_\odot$)\\
       \hline \noalign{\smallskip}
       2MASS\,J04202144+2813491&\tauZero&04 20 21.4 &+28 13 49.2& Taurus & M1 & \\
       \hh & \hh & 04 31 37.5 & +18 12 24.5& Taurus & M0\\
       IRAS\,04302+2247& \IRAS &  	04 33 16.5 &+22 53 20.4& Taurus & K6-M3.5 & $1.3 - 1.7$\\
       \hk & \hk & 04 31 50.6 &+24 24 16.4 & Taurus &  M0.5 & $0.89\pm0.04$ \\
       \hv & \hv & 04 38 35.5 & +26 10 41.3 & Taurus & K6 & $1.33\pm 0.04$\\
       IRAS\,04200+2759 &\JTroisZeroSept&04 23 07.8 & +28 05 57.5& Taurus & M3.5-M6&$0.52 \pm 0.04$\\ 
       FS\,Tau\,B& \JDeuxDeux &	04 22 00.7 &+26 57 32.5 & Taurus & K5\\
       IRAS\,04158+2805& \irasGerrit~&04 18 58.1 &+28 12 23.4& Taurus & M5.25\\
       2MASS\,J16313124-2426281& \Oph & 16 31 31.3 & -24 26 28.5& Ophiuchus & K4-K5$^d$ & $1.2 \pm 0.2^d$\\
       \Junun & \Junun& 11 11 10.8 & -76 41 57.4& Cha\,I & M2.5\\
       \Junsois& \Junsois& 11 16 02.8 & -76 24 53.2& Cha\,I&K8\\
        \hhq & \hhq& 11 04 22.8 & -77 18 08.0&Cha\,I&K7\\
        \hline
    \end{tabular}
    \tablefoot{Coordinates are J2000. SFR: Star-forming region, SpT: Spectral type, M$_\star$: Stellar masses, from dynamical estimates based on gas emission.}
    \tablebib{$^{(a)}$ \citet{Luhman_2007}, $^{(b)}$ \citet{Luhman_2010}, $^{(c)}$~\citet{Simon_2019},  $^{(d)}$~Flores et al. (in prep)}
    \label{tab:stellar_parameters}
\end{table*}

\subsection{ALMA edge-on survey}
Our ALMA cycle 4 and 5 observations (Project 2016.1.00460.S, PI: M\'enard) were divided into two groups, targeting respectively seven sources located in Taurus and three in Chameleon\,I.  The observations were performed in band~7~(0.89\,mm) and in band~4~(2.06\,mm). We present the different setups in the following section. 

\subsubsection{Band~7 survey}
\label{sec:b7_reduction}
The band~7 observations of the Taurus sources (\tauZero, \IRAS, \hk, \hv, \JTroisZeroSept, \JDeuxDeux, and \irasGerrit) were performed with both a compact and an extended array configuration. For the Chamaeleon sources (\Junun, \Junsois, and \hhq), only the compact configuration was observed. The observational setup is summarized in Table~\ref{tab:obsEOD}.  The spectral setup was divided into three continuum spectral windows, with dual polarization, 2~GHz bandwidth spectral windows centered at~344.5, 334.0, and 332.0\,GHz, and one spectral window set to observe the~$^{12}$CO J=$3-2$ transition at~345.796\,GHz.  In this paper, we focus on the continuum data which has a geometric mean frequency of~336.8\,GHz~(0.89\,mm). The reduction and analysis of the~CO data will be presented in a separate study. Because the observations were performed over two cycles, different versions of CASA have been used for the calibration.

\begin{table*}[]
    \centering
	\small
	 \caption{ALMA observations.}
  	  \begin{tabular}{lcclccccc}
	  \hline\hline
    	Source& Band & Obs. date & Config.& Baselines & N$_{ant}$ &t$_{int}$ (min)&v$_\mathrm{CASA}$& Project ID  \\
    	~~\,(1)& (2)& (3)& ~~\,(4)& (5)& (6)& (7)& (8)& (9)\\
    	\hline
	Survey Tau$^a$ &7  & 24/11/2016& C40-4 & 15 m$-$0.7 km &43 & 1.1  & 4.7 &2016.1.00460.S\\
	  &  & 18/08/2017& C40-7& 21 m$-$3.6 km &43 & 3.7  & 5.1 &2016.1.00460.S\\
	  &4   & 27/09/2017& C40-8/9& 41 m$-$14.9 km &43 & 6.7  & 5.1 &2016.1.00460.S\\ 
	Survey Cha$^b$  &7   & 15/11/2016& C40-4/6& 15 m$-$0.9 km &42 & 1.1  & 4.7&2016.1.00460.S \\
	\hh  &7 &  14, 15,  21/10/2016& C40-6& 19 m$-$2.5 km &42-46 & 174  & 4.7&2016.1.01505.S\\
	 &6 &  19, 21/07/2015& C34-6/7& 13 m$-$1.6 km&42 & 78  & 4.3  &2013.1.01175.S \\
	  &4  & 23/10/2016& C40-6 & 19 m$-$1.8 km &48 & 12  & 4.7&2016.1.01505.S\\
	 &  & 12/09/2017& C40-8/9& 41 m$-$ 9.5 km &44 & 31  & 4.7&2016.1.01505.S\\
	\tauZero &6  & 05/12/2016& C40-3 & 15 m$-$ 0.7 km &41 & 7.5  & 4.7 &2016.1.00771.S\\
	 & & 21/10/2016&C40-6& 18 m$-$ 1.8 km &44 & 25  & 4.7 &2016.1.00771.S\\
	\Oph &6  & 25/04/2017& C40-3 & 15 m$-$0.5 km &41 & 8.5  & 4.7 &2016.1.00771.S\\
	 & & 07/07/2017& C40-5 &17 m$-$ 2.6 km &44 & 27  & 4.7 &2016.1.00771.S\\
	\hline
	 \end{tabular}
	 \tablefoot{Column 1: Target name, Column 2: Observing band, Column 3: Observing date, Column 4: ALMA array configuration, Column 5: Minimum and maximum baseline range, Column 6: Number of antennas, Column 7: On source integrating time, Column 8: Calibrating CASA version. $^{(a)}$ The Taurus sources included in the survey are \tauZero, \IRAS, \hk, \hv, \JTroisZeroSept, \JDeuxDeux, and \irasGerrit. $^{(b)}$  The Chamaeleon\,I sources included in the  survey are \Junun, \Junsois, and \hhq.}
	 \label{tab:obsEOD}
 \end{table*}

We calibrated the raw data of the compact array executions using the pipeline from CASA package version~4.7. The raw data of the extended configuration observations of the Taurus sources was manually calibrated, using CASA version~5.1. 
Whenever possible we used the supplied ALMA phase calibrator in the dataset~(for \tauZero, \JDeuxDeux, \irasGerrit). However, for four targets~(namely \hk, \hv, \JTroisZeroSept, and \IRAS), the spectral window setup used for the supplied phase calibrator was different from that of the science target. These data could not be reduced using the standard pipeline.  For \hv~and \JTroisZeroSept, it was possible to use the check source as a phase calibrator, and for \hk, the more distant bandpass calibrator could be used. In these cases, the calibrator was only observed once before each of the science targets (rather than bracketing it in time); this increased the phase calibration uncertainties for these objects.
\IRAS~did not have a usable phase calibrator. However, this source is bright and extended, and self-calibration using the average of all spectral windows could be performed without an initial phase reference. Consequently for this target, there was no absolute astrometric solution in the extended configuration data;  also the achievable angular resolution was somewhat worse than the other sources of the sample.  

\subsubsection{Band~4 survey}
The band~4 observations of the edge-on survey were only performed for the sources located in Taurus (\tauZero, \IRAS, \hk, \hv, \JTroisZeroSept, \JDeuxDeux, and \irasGerrit).  
The continuum spectral windows were centered on 138, 140, 150, and 152\,GHz, with a geometric mean frequency of 145.0\,GHz~(2.06\,mm). The relevant parameters of the observations are reported in Table~\ref{tab:obsEOD}. 
The raw data were pipeline calibrated using the CASA package, version~5.1. 

\subsection{Archival and literature data}
\label{sec:archival_data}
\subsubsection{ALMA observations of \hh}

We include multi-wavelength, band~4,~6, and~7, observations of \hh~in the present study. The observational setup and data reduction of the band~6 observations are presented in \citet[][Project 2013.1.01175.S, PI: Dougados]{Louvet_2018}. 
We also use new band~4 and band~7 cycle 4 observations~(Project 2016.1.01505.S, PI: Louvet), and present the data reduction in the following paragraphs.\\

The band~7 observations of HH~30 were performed with only one array configuration, in four executions between October~14 and October~21, 2016~(see Table~\ref{tab:obsEOD}). 
The dual polarization spectral setup included two 2~GHz bandwidth spectral windows for the continuum emission, centered at 331.6 and 344.8 GHz, and two other spectral windows set to observe the $^{13}$CO and C$^{18}$O J=3-2 emission. Here we present only the continuum observations. 
The observations were calibrated by the ALMA pipeline using CASA~4.7.\\ 

The band~4 observations were performed with two configurations: a compact configuration and an extended configuration~(see Table~\ref{tab:obsEOD}). 
The ALMA correlator was configured to record dual polarization with four separate continuum spectral windows of 2 GHz each centered at 138, 140, 150, and 152\,GHz. 
The observations were calibrated by the pipeline using~CASA~4.7.

\subsubsection{ALMA observations of \tauZero~and \Oph}
We also include band~6 cycle~4 observations of~\tauZero\:and \Oph~(Project 2016.1.00771.S, PI:~Duch\^{e}ne). Although the spectral setup targeted emission lines of three CO isotopologues, we focus here on the continuum data. An analysis of the line emission of \Oph\:will be presented in a separate study~(Flores et al., in prep). 

For both sources, the two continuum spectral windows were centered on~216.5 and~232.3\,GHz leading to a geometric mean frequency of~224.4\,GHz~(1.34\,mm). The detailed observational setup is presented in Table~\ref{tab:obsEOD}.  
Data from both configurations were reduced using the pipeline from CASA package version~4.7.

\subsection{ALMA imaging}
\label{sec:ALMAimaging}
\begin{figure*}
    \centering
    \includegraphics[width =0.95\textwidth]{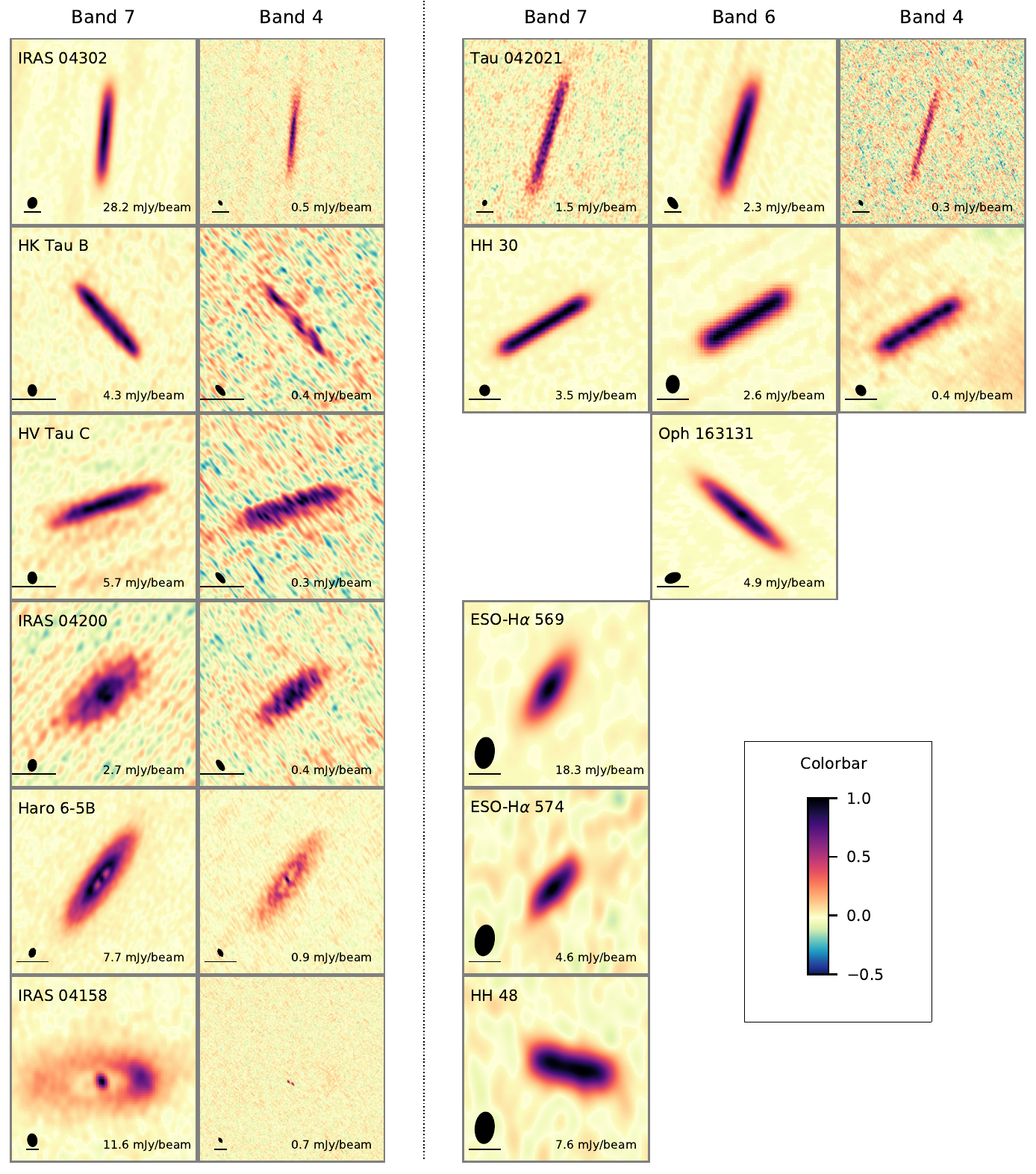}
     \caption{Images of all sources included in this study normalized to their peak intensity~(reported in the bottom right corner of each image). Each column corresponds to a different band and each line displays two sources, separated by the vertical dashed line. We show the beam size (ellipse) and a 0.5\arcsec\:scale (dark line) in the bottom left corner of each panel.}
    \label{fig:alma_images}
 \end{figure*}

We constructed the images from the calibrated visibilities with a Briggs robust weighting parameter of 0.5, except for the band~6 of \hh\: for which we used the reduction presented in \citet[][robust of 0.56]{Louvet_2018}. When the data were taken with several array configurations~(see Table~\ref{tab:obsEOD}), the images were generated by combining the visibilities from the compact and extended configurations. The only exception is \irasGerrit\: for which we show the band~7 compact configuration observations only. We applied self-calibration on the sources with the highest signal-to-noise ratio~(on at least one array configuration). In band~7 this corresponds to \tauZero, \IRAS, \hk, and \JDeuxDeux, in band~6 to \tauZero~and \Oph, and to only \hh~in band~4. For the Taurus survey, this leads to typical beam sizes of about~0.11\arcsec$\times$0.07\arcsec\:in band~7 and about~0.11\arcsec$\times$0.04\arcsec\:in band~4. For the other sources, the beam sizes range from 0.23$\times$0.13\arcsec\:to 0.48$\times$0.30\arcsec. The beam sizes of each observations are reported in Table~\ref{tab:beams}, and we present the images in Fig.~\ref{fig:alma_images}.

Additionally, we also recompute the maps to get a unique angular resolution for each source observed both in band~4 and band~7 (except for \irasGerrit, for which the disk is not detected in band~4). To do so, we first re-imaged each source using the same \texttt{uvrange} and applying a \texttt{uvtaper} to limit the effect of flux filtering and of different uvcoverage. Then, to obtain exactly the same angular resolution between both bands, we used the \texttt{imsmooth} CASA task. The restored beam sizes are reported in the last column of Table~\ref{tab:beams}.

\subsection{Astrometric accuracy and map registration}
ALMA maps are subject to astrometric uncertainties due to limited signal-to-noise on the phase calibrator and errors in the phase referenced observations related to weather or antenna position errors. To compute accurate spectral index maps~(see Section~\ref{sec:alpha_maps}), the images at the different wavelengths have to be accurately aligned. In this section, we discuss the registration of our images.

For each source and band, we performed simple Gaussian fits in the image plane to estimate the centroid position. The offsets between band~4 and band~7 ranged from 2\,mas~(milliarcseconds) for~\tauZero\:and \JTroisZeroSept, up to 60\,-\,90\,mas for~\hh, \hv, and \IRAS. Most are larger than expected from source proper motion between the different executions and from the astrometric accuracy of ALMA~(10\% of the synthesized beam, see ALMA Technical Notebook). 
However, for the three disks with the largest offsets~(\hh, \hv, and \IRAS), different phase calibrators were used between the observations. In addition, the standard phase calibration could not be performed for the band~7 data of \IRAS~and \hv~(see above), and the weather conditions during the band~7 extended configuration observations were relatively poor, with phase rms of $\sim 64^\circ$ on the longest baselines. This will also deteriorate the positional accuracy.  
Overall, the astrometric accuracy of our observations is not sufficient to confirm any significant motion or shift of the sources between the two bands. Because models of edge-on disks also show that offsets between bands should remain minimal compared to the beam size, we registered our images to a common center in both bands. We aligned the band~4 and band~7 observations using the \texttt{fixplanets} task in CASA\footnote{We note that we set the option \texttt{fixuvw} to True when applying the \texttt{fixplanets} task, which is similar to using the \texttt{fixvis} task.}. \\

\section{Results}
\label{sec:results}

\subsection{Continuum emission and brightness temperatures}
\label{sec:images}

\begin{table}[]
    \centering
    \caption{Millimeter fluxes.}
    \begin{tabular}{ccccc}
    \hline
    \hline
    Sources & F$_\mathrm{B7}$ (mJy)& F$_\mathrm{B6}$ (mJy)& F$_\mathrm{B4}$ (mJy)\\
    \hline
    \tauZero& $124.2\pm12.4$& $47.2 \pm4.7$&$15.4\pm1.5$\\
    \hh& $54.5\pm5.5$&$~22.3\pm0.1^a$& $6.9\pm0.7$\\
    \IRAS& $267.5 \pm 26.8$&&$37.2\pm3.7$ \\
    \hk &$55.6 \pm 5.6$ && $4.4\pm 0.4$ \\
    \hv&$90.6\pm9.1$&&$11.6\pm1.2$\\
    \JTroisZeroSept& $65.6\pm6.6$&&$11.6\pm1.2$ \\
    \JDeuxDeux&  $340.9\pm34.1$&&$35.3\pm3.5$\\
    \irasGerrit&$259.2\pm25.9^*$&& $2.0 \pm 0.2^*$\\
    \Oph&$125.8 \pm 2.4^b$&$44.8\pm 4.5$ &\\
    \Junun&$40.2\pm4.0$&& \\
    \Junsois&$~9.3\pm0.9$&& \\
    \hhq& $31.0\pm3.1$&&\\
    \hline
    \end{tabular}
    \tablefoot{Total fluxes are measured using elliptical apertures centred on the targets. $^{(*)}$ Band 7: Total flux for the disk and central binary. Band 4: Only flux from the central binary.}
    \tablebib{$^{(a)}$ \citet{Louvet_2018}, $^{(b)}$ \citet{Cox_2017}.}
    \label{tab:results_fluxes}
\end{table}

All millimeter-wavelength continuum images are presented in Fig.~\ref{fig:alma_images}. 
The majority of the disks in our sample show an elongated emission shape, with large axis ratios and in several cases roughly constant surface brightness along the major axis, confirming that they are highly inclined.  

Two disks in the sample, however, present a different shape. \JDeuxDeux~and \irasGerrit\: show the presence of a ring and central emission peak. We evaluate the position of the rings in Section~\ref{sec:radial_extent}. As in previous CARMA observations~\citep{Sheehan_2017}, the band~7 image of \irasGerrit\:reveals a highly asymmetric ring (brighter on the western side) with a compact source toward the center.  This ring is not detected in our band~4 observations, mostly due 
to flux dilution into small beams. However, the band~4 observations clearly resolve the emission at the center of the ring into two point sources. \irasGerrit\:is a binary system. A detailed analysis of this source will be presented in a dedicated study (Ragusa et al., in prep). 

The total continuum flux density of each source is measured by integrating the signal, down to the 3$\sigma$ noise level, within elliptical apertures tailored to each source. The flux densities are reported in~Table~\ref{tab:results_fluxes}. 
We use 10\% error values throughout Table~\ref{tab:results_fluxes}, which correspond to the typical flux calibration errors of ALMA~(see ALMA Technical Handbook\footnote{https://almascience.eso.org/documents-and-tools/latest/documents-and-tools/cycle8/alma-technical-handbook}). 
We note that although the phase calibration method used for the extended configuration observations in band~7 was non-standard~(see Section~\ref{sec:b7_reduction}), the flux calibration followed the usual procedure and, accordingly, the flux calibration uncertainty should be nominal. We checked that the integrated fluxes recovered using the compact configuration band 7 observations or both compact and extended configurations jointly are consistent within 10\%.\\

To further investigate the disk properties, we calculate the brightness temperature, $T_B$, maps of each source. To ease the comparison between bands, we estimate the brightness temperatures from the band~7 and band~4 maps computed at the same angular resolution (see Section~\ref{sec:ALMAimaging}). 
We do not use the Rayleigh-Jeans approximation. We show the brightness temperature profiles measured along the major axis in~Fig.~\ref{fig:brightness_T} and report the \emph{peak} brightness temperatures in Table~\ref{tab:brightness_T}. 

We note that, even for the most resolved sources~(i.e., least impacted by beam dilution), the inferred brightness temperatures are lower or comparable to those required to be in the Rayleigh-Jeans regime, respectively T\,$>$\,16.2\,K for band~7 and~6.8\,K for band~4. This strengthens our choice not to use this approximation. 
Additionally, we find that the band~4 brightness temperatures are systematically lower than the band~7 ones.

\begin{table}[]
    \centering
    \caption{Peak brightness temperatures.}
    \begin{tabular}{cccc}
        \hline\hline
         Sources &  B7 (K)&   B4 (K)\\
         \hline
         \tauZero&$6.9^\dagger$&$5.3^\dagger$\\
         \hh &$7.2~$& $3.7~$\\ 
         \IRAS &$10.3~$&$6.7^\dagger$ \\
         \hk&$11.4~$ &$5.4^\dagger$ \\
         \hv & $13.5~$&$6.8$\\
         \JTroisZeroSept& $~9.1^\dagger$&$8.7^\dagger$\\
         \JDeuxDeux& $15.8^\dagger$&$9.9^\dagger$ \\
         \irasGerrit&$5.2~$& \\ 
         \Junun&$6.5~$&\\
         \Junsois&$4.3~$&\\
         \hhq& $4.9~$&\\
        \hline
    \end{tabular}
    \tablefoot{The uncertainties are limited by errors in flux calibration and can thus be estimated by 10\% of the reported brightness temperature peak. The reported peak brightness temperatures were computed using the same beam in both bands (see Table~\ref{tab:beams}). We only report the peak band~7 brightness temperature of \irasGerrit\:because the disk is not detected in band~4. $^{(\dagger)}$ Well resolved sources in all directions.}
    \label{tab:brightness_T}
\end{table}

\begin{figure*}%[h]
    \centering
    \includegraphics[width =0.87\textwidth]{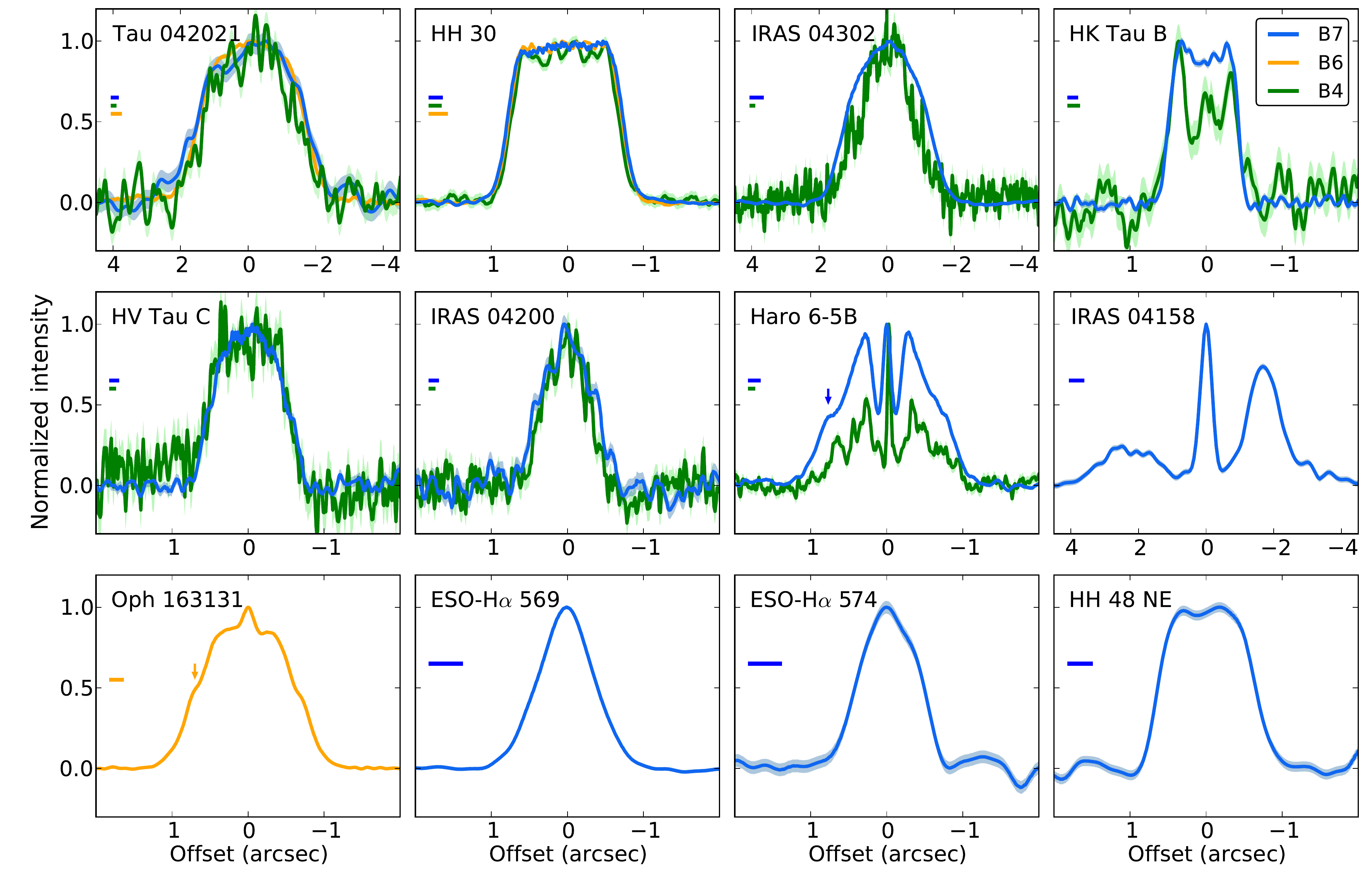}
    \caption{Normalized major axis intensity profiles. Data in the three bands are represented by the green~(band~4), orange~(band~6) and blue~(band~7) lines. The light shading corresponds to the normalized rms in each band. The beam sizes in the direction of the cut are shown in the left part of each plot as horizontal lines. We indicate the shoulders of \JDeuxDeux~and \Oph~by an arrow. We smoothed the cuts through \tauZero~by convolving them by a 1-D Gaussian of the beam width, to reduce the noise and make the plot clearer.}
    \label{fig:radial_cuts_maj}
    \bigskip 

    \centering
    \includegraphics[width =0.87\textwidth]{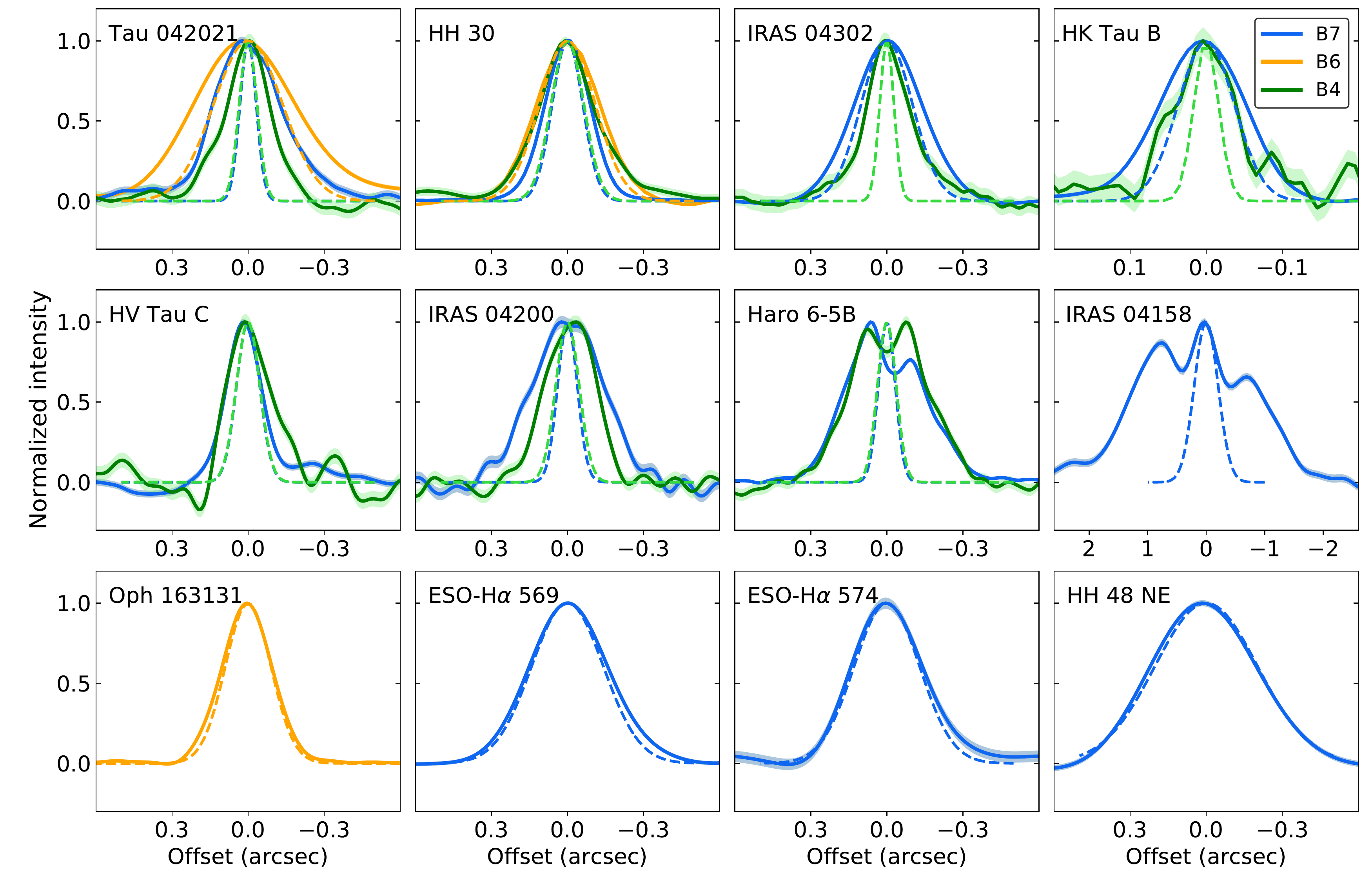}
    \caption{Normalized averaged minor axis intensity profiles. For the most inclined sources (resp. least inclined: \JTroisZeroSept, \JDeuxDeux, and \irasGerrit), these were obtained by averaging minor axis intensity cuts over the whole major axis extent of the disk (resp. over the central 0.3\arcsec). The light shading correspond to the normalized rms in each band. The beam sizes in the direction of the minor axis are shown as dotted Gaussian of the corresponding color.}
    \label{fig:radial_cuts_min}
\end{figure*}

\begin{table*}[]
    \centering
    \caption{Position angle, inclination, major axis sizes, and estimates of the mean deconvolved major axis sharpness of the millimeter data.}
    \begin{tabular}{crc||ccc||c||c}
        \hline \hline
         Sources & PA & $i_\mathrm{AxisRatio}$ & B7 & B6 & B4& Optical-NIR & $\Delta r / r$\\
         &(\degr)& (\degr) & (\arcsec) & (\arcsec) & (\arcsec) & (\arcsec) &\\
         \hline
         \tauZero&-16 & $>85$ & $4.10 \pm 0.01$& $3.76 \pm 0.02$& $3.67 \pm 0.01$& 5.0&$0.6 \pm0.2$\\
         \hh & 121 &$~>85^a$& $1.85\pm0.01$ & $1.82 \pm 0.02$& $1.73\pm 0.01$&3.1 & $0.2 \pm 0.1$\\
         \IRAS& 175 & $>84 $& $3.15 \pm 0.03$ & & $2.84 \pm 0.01$ & &$0.7 \pm 0.1$\\
         \hk&41& $>83 $ &$0.99 \pm 0.01$&&$0.99 \pm 0.01$& 1.3& $0.2 \pm 0.1$\\
         \hv & 108 & $>80 $& $1.20 \pm 0.01$ & & $1.18 \pm 0.01$& 0.8 & $0.5\pm0.2$\\
         \JTroisZeroSept& 129 & $69 \pm 2$ & $1.00 \pm 0.01$ &&$0.85 \pm 0.01$&  & $1.0 \pm 0.4$\\
         \JDeuxDeux & 145 & $74 \pm 2$ & $2.06 \pm0.01$&& $1.74 \pm 0.01$&2.3& $~0.8 \pm 0.3^\dagger$\\
         \irasGerrit&92& $62 \pm 3$  &$7.47\pm 0.03$&&&13.6&\\
         \Oph &49 & $> 80$& &$2.50 \pm 0.01$&&2.6&$0.7 \pm 0.2$\\
         \Junun&144&$> 64$&$1.88 \pm 0.04$&&&2.0&$0.9 \pm 0.2$\\
         \Junsois&141& $>69$ &$ 1.35 \pm 0.04$&&&1.2&$0.7\pm0.2$\\ 
         \hhq& 75& $>68$& $1.72\pm 0.03$&&& 1.3& $0.4\pm0.1$\\
        \hline
    \end{tabular}
    \tablefoot{For each source, the errors in the millimeter sizes correspond to one tenth of the beam width in the major axis direction. When possible, the $\Delta r/r$ ratios correspond to their averaged values between the band~7 and band~4. All are deconvolved by the beam. We also indicate an estimate of the inclination inferred from the millimetric mean axis ratio. These are lower limits when the disks are not resolved along their minor axis in all bands by more than 2 beams, or when the measured inclination is too high not to be influenced by the physical vertical extent of the disk. $^{(\dagger)} \Delta r/r$ of the band~7 only.}
    \tablebib{$^{(a)}$ \citet{Louvet_2018}}
    \label{tab:major_axis_cut}
\end{table*}

\subsection{Surface brightness profiles}
\subsubsection{Radial extent}
\label{sec:radial_extent}

To characterize the radial extent of the disks, we present cuts 
along the major axis of each disk in Fig.~\ref{fig:radial_cuts_maj}. The cuts are normalized to their maximum intensity. For \tauZero, \IRAS\:and \hv, for which the band~4 emission is very noisy, we estimate the normalizing factor as the amplitude of the Gaussian best fitting the curves. All sources are well resolved along their major axis. 

\paragraph{Morphology.} For several sources, the brightness profile is flat along the major axis direction and drops steeply at the edges. This is particularly true for \hh, \hk, \hv, and \hhq. The lack of a central brightness peak further supports the idea that these disks are optically thick and highly inclined, so that we are viewing only the outer optically-thick edge. 
Conversely, the disks of \tauZero, \IRAS, \JTroisZeroSept, \Junun, and \Junsois\:show more centrally peaked emission without a clear plateau, suggesting that they are optically thinner and/or viewed with a lower inclination, less edge-on.
We note that the radial brightness profile of  \hk~shows hints of rising at the edges. While this is a marginal (3$\sigma$) feature, this may indicate the presence of a ring~(or transition disk).

Finally, three disks (\JDeuxDeux, \irasGerrit, and \Oph) show symmetric shoulders or clear evidence of ring like features.
The main ring of \JDeuxDeux~peaks at~$\sim$0.29\arcsec~(41\,au) in both bands. Furthermore, a shoulder seen in band~7 is associated with a peak in the higher-resolution band~4 cut; this would correspond to a fainter ring located at~$\sim$0.77\arcsec~(108\,au). 
The brightness asymmetry of the outer ring in \irasGerrit~is very clear in the major-axis cut: its western side is about~3~times brighter than the eastern side. We also note that the ring is slightly off-centered compared to the central binary. The western side of the disk peaks at~$\sim$1.71\arcsec~(239\,au), while the eastern side peaks at~$\sim$2.28\arcsec~(319\,au) from the center. We fit Gaussians on each side of the disk in the radial profile and find that the eastern ring is about 1.7 times wider than the western ring~(with full width half maximum, FWHM, of~1.98\arcsec\,and~1.18\arcsec\:respectively). 
We also find that \Oph~displays a relatively flat profile in the inner~0.5\arcsec, but has a sharp central peak. It has symmetric shoulders at $\sim$0.70\arcsec\, radius, which suggests that this disk contains two rings and is viewed slightly away from edge-on. Further modeling of this source, focusing on dust emission, will be presented in a separate analysis~(Wolff et al., in prep).

\paragraph{Size estimations.} 
To determine the sizes, we first normalized the major axis cuts presented in Fig.~\ref{fig:radial_cuts_maj}, as mentioned in the beginning of this section. 
Then, for each source, we estimate the relevant 3$\sigma$ noise level in the image with the worst signal-to-noise ratio, either band~4 or band~7. This 3$\sigma$ level is converted into a fraction of the peak, in percent. The sizes in both bands are then measured at the same level, which means at the same fraction of the peak. The size of the disks along their major axis are reported in Table~\ref{tab:major_axis_cut}. The errors correspond to a tenth of the beam size along the major axis direction. 
We verified that using the FWHM of the cut profiles instead of the 3$\sigma$ levels yields similar conclusions. This is also the case for the cuts obtained from the maps computed with a similar uv-coverage.

Along with the major axis size, we also estimate the apparent sharpness of the disk edges, as measured in the image plane. To do so, we measure the fractional range of radius over which the millimeter emission decreases from~80\% to~20\% of the peak emission, called $\Delta r / r$. 
We define $\Delta r$ by $\Delta r = |r_\mathrm{80\%} - r_\mathrm{20\%}|$ and the normalization radius by  $r = |r_\mathrm{80\%} + r_\mathrm{20\%}| / 2$. We chose this flux range because the radial profiles can usually be well approximated by straight lines in this interval.
We report the values of $\Delta r/r$ in the last column of Table~\ref{tab:major_axis_cut}. The values are the mean of the estimations in band~4 and band~7, and they are deconvolved by the beam size. Sharp outer edges have small $\Delta r$ and hence small $\Delta r/r$. 
For example, a typical beam with a FWHM of 0.1\arcsec\:would have $\Delta r/r\simeq0.1$ when calculated in a small disk similar to \hk~($r = 0.5\arcsec$), or $\Delta r/r\simeq0.04$ in a disk as large as \IRAS~($r=1.2\arcsec$).
All the disks in our sample have $\Delta r/r$ values larger than 0.2, which corresponds to shallower outer edges than the typical beam.\\

We find two different families of objects when comparing the radial extent of the disks in band~4 and band~7. Four sources (\tauZero, \IRAS, \JTroisZeroSept, and \JDeuxDeux) show a band~7 size more extended than the band~4 by more than 2 beams.
When compared with the same angular resolution and uv-coverage, these sources have band~7 major axis size on average~12\% larger than that of the band~4. As the band~7 shorter wavelength traces smaller particles than those probed by band~4, the smaller sizes in band~4 suggest that the larger particles have drifted inward relative to the smaller ones. The outer edges of these four sources are well resolved and they have an average apparent sharpness of~$\Delta r/r \sim 0.8$, much shallower than the beam.

For the three remaining sources~(\hh, \hk, \hv), no difference in radial extent is found between band~7 and band~4.  This might suggest the presence of dust traps at the outer edges of these disks, which can slow radial drift and lead particles to stop at particular radial locations~\citep{Powell_2019, Long_2020}. 
Including \hhq, these four sources have the sharpest edges, with $\Delta r/r$ between~0.2 and~0.5. The edges of these disks are only marginally shallower than the typical beam.  We note that 3 out of these 4 systems are known binaries and dynamical interactions may also lead to sharp outer edges.

\begin{table*}[htb]
    \centering
     \caption{Full width half maximum of the (averaged) minor axis profiles, measured and deconvolved by the beam size.}
    \begin{tabular}{cccc|ccc}
        \hline
        \hline
         Sources & Minor B7 & Minor B6& Minor B4&Deconvolved B7 & Deconvolved B6& Deconvolved B4\\
          & (\arcsec) & (\arcsec) & (\arcsec) & (\arcsec) & (\arcsec) & (\arcsec)\\
         \hline
         \tauZero& $0.32 \pm 0.01^\dagger$ & $0.49 \pm 0.03$ & $0.21 \pm 0.01^\dagger$&$0.31 \pm 0.01^\dagger$&$0.37 \pm 0.06$& $0.20 \pm0.01^\dagger$\\
         \hh& $0.21 \pm 0.01~$ & $0.29 \pm 0.03$ & $0.27 \pm 0.02~$ & $ 0.16 \pm 0.02~$ & $0.14 \pm 0.07$& $0.22\pm 0.02~$ \\
         \IRAS& $0.32 \pm 0.02~$ && $0.21 \pm 0.01^\dagger$ &$0.21 \pm 0.05~$&& $0.20 \pm 0.01^\dagger$\\
        \hk& $0.13 \pm 0.01~$ && $0.10 \pm 0.01^\dagger$& $0.10 \pm 0.01~$ && $ 0.09 \pm 0.01^\dagger$\\
         \hv &$0.17 \pm 0.01~$ & & $0.19 \pm 0.01$ & $ 0.13 \pm 0.02~$ && $0.16\pm 0.02$\\
         \JTroisZeroSept$^*$& $0.33 \pm 0.01^\dagger$ & & $0.23\pm 0.01^\dagger$& $ 0.31 \pm 0.01^\dagger$ && $ 0.21 \pm 0.01^\dagger$ \\
         \JDeuxDeux$^*$ & $0.35 \pm 0.01^\dagger$&&$0.35 \pm 0.01^\dagger$ & $0.34 \pm 0.01^\dagger$&&$ 0.34 \pm0.01^\dagger$ \\
         \irasGerrit$^*$&$2.36 \pm 0.05^\dagger$&&&$ 2.31\pm 0.05^\dagger $&&\\
         \Oph&$0.16 \pm 0.07^a$ &$0.23 \pm 0.02$&&&$ <0.14$&\\
         \Junun & $0.37 \pm 0.03~$&&&$<0.13$&& \\
         \Junsois& $0.33 \pm 0.03~$ &&&$<0.21$&&\\
         \hhq & $0.49 \pm 0.05~$&&& $<0.24$&&\\
         \hline
      \end{tabular}
    \tablefoot{The uncertainties on the measured minor axis sizes correspond to one tenth of the beam size along the cut direction. $^{(*)}$~Disks with inclination lower than 75$^\circ$. Their minor axis size is likely dominated by the radial extent of the disk over its vertical extent, as opposed to more inclined disks. $^{(\dagger)}$~Resolved by more than 2 beams in the minor axis direction. These are the ones for which the deconvolved minor axis size should be the most reliable.}
    \tablebib{$^{(a)}$ \citet{Cox_2017}}
    \label{tab:radial_cuts_min}
\end{table*}

\subsubsection{Disk extent perpendicular to the midplane}
\label{sec:vertical_extent}

For the most inclined systems, the brightness maps shown in Fig.~\ref{fig:alma_images} have very elongated, linear shapes rather than elliptical ones, so we generate the minor axis profiles by taking the mean of the cuts at all distances along the major axis. In that case, the size of the minor axis is dominated by the vertical extent of the disk perpendicular to the midplane. For the less inclined sources where a clear ellipticity is visible in the image~(namely \JTroisZeroSept, \JDeuxDeux, and \irasGerrit), we generate the minor axis profiles by averaging over a restricted range, only $\pm0.15$\arcsec\: around the center of the disk. In that case, the minor axis is dominated by the projection of the disk radius. 
We show the averaged brightness profile along the minor axis for all sources in Fig.~\ref{fig:radial_cuts_min}. 
Dashed lines trace the Gaussian beam profiles along the cut direction. 

We find that six out of the twelve disks of our sample are well-resolved along the minor axis, having a width more than twice the beam width. They are \tauZero, \JTroisZeroSept, \JDeuxDeux~(in both band~7 and band~4), \irasGerrit~(in band~7), and \IRAS, \hk~(in band~4). Additionally, the minor axis profiles of \JDeuxDeux~and \irasGerrit~reveal the presence of rings, and we see a clear asymmetry in the band~7 cut of \JDeuxDeux.

We measure the minor axis sizes by fitting Gaussians to the generated profiles, Gaussians being good first approximations.
For \JDeuxDeux~and \irasGerrit~for which the cuts show ring features, we estimate the FWHM directly through the cuts, without fitting a Gaussian. We report the resulting FWHM in Table~\ref{tab:radial_cuts_min}, where the errors correspond to a tenth of the beam size projected in the direction of the cut\footnote{We note that even for the most inclined disks, the minor axis size is related but is not a measurement of an equivalent ''dust scale height'' at a given radius. Indeed, observations are measuring the integrated intensity along the line-of-sight over the whole disk~(i.e., from several radii) and are affected by optical depth effects.}. 

In order to extract the intrinsic vertical extent of the disks, we also deconvolve the minor axis sizes by the beam, assuming that both profiles are Gaussian. These values are presented in the last columns of~Table~\ref{tab:radial_cuts_min}. 
For the least resolved disks (i.e., for \Oph, \Junun, \Junsois, and \hhq), we only report upper limits. We estimate the uncertainties on the deconvolved minor axis sizes by propagating the errors, assuming that the error on the beam size is 10\% of the beam major axis size.\\ 

We also estimate the disk inclinations from the measured axis ratio in all bands, and report them in Table~\ref{tab:major_axis_cut}. For consistency, we use the FWHM of the major axis sizes~(cut at 50\% of the peak flux) to estimate the inclination, as opposed to the size at 3$\sigma$ reported in Table~\ref{tab:major_axis_cut}. 
For the sources that are not resolved in the vertical direction~(\Oph, \Junun, \Junsois, \hhq) and those with the smallest axis ratio~(\tauZero, \IRAS, \hk, \hv), we present these values as lower limits. 
Indeed, for the most inclined systems (i.e. those with the smallest axis ratio), the minor axis size might not be dominated by the inclination but by the actual vertical thickness of the disks, which leads to lower apparent inclinations based on the axis ratio only.
Except \JTroisZeroSept, \JDeuxDeux, and \irasGerrit, all resolved disks have an inclination larger than~75$^\circ$. We note that \tauZero~is the only highly inclined disk resolved along its minor axis in both band~7 and band~4. For this disk, the band~7 appears about 1.5 times more extended vertically than the band~4.

\begin{figure*}
    \centering
    \includegraphics[width =\textwidth]{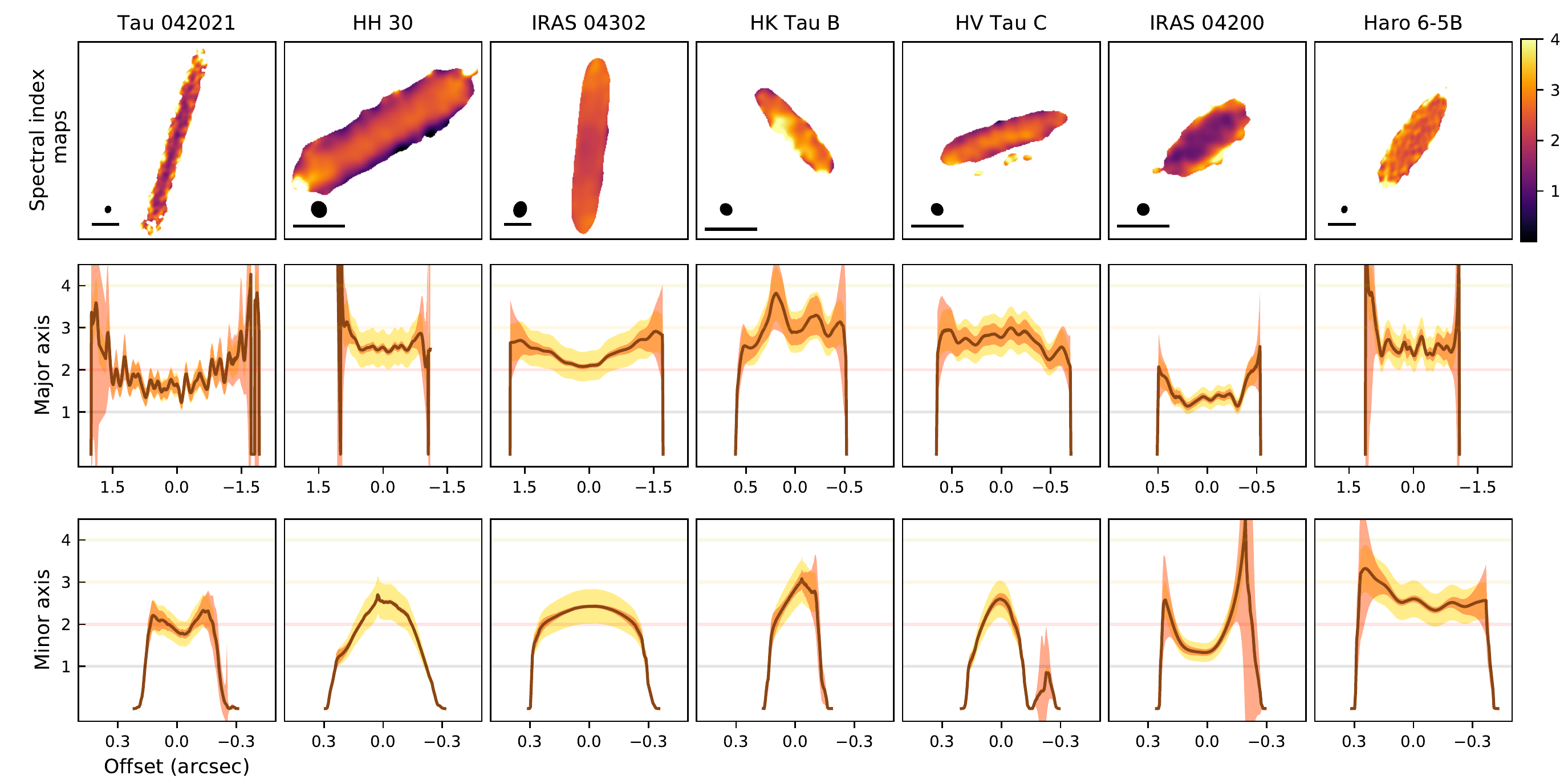}
    \caption{\emph{Top row:} Spectral index maps, applying a filter to keep only the pixels with more emission than 5$\sigma$ in both bands. The beam size is shown in the bottom left corner, along with a dark line representing a 0.5\arcsec\:scale. \emph{Middle row:} Spectral index cuts along the major axis. \emph{Bottom row:} Spectral index profiles along the minor axis, averaged as done for Fig.~\ref{fig:radial_cuts_min} (see text Section~\ref{sec:vertical_extent}). For all cuts, yellow errors correspond to a flux calibration error of 10\% in both bands, while orange errors are estimated from the signal-to-noise in each band. The x-axis corresponds to the offset to the center of the disk in arcseconds. }
    \label{fig:alpha_map_cuts}
\end{figure*}

\subsection{Estimation of spectral indices}
\label{sec:alpha_maps} 
The millimeter spectral index, defined as $F_\nu \propto \nu^{\alpha_{mm}}$, can be used to study optical depth and grain growth in a disk~\citep[see e.g., review by][]{Williams_Cieza_2011}. 
Indeed, assuming scattering is negligible, the millimeter intensity can be expressed as~$I_\nu~=~B_\nu(T)(1~-~e^{-\tau_\nu})$, in which~$B_\nu(T)$ is the Planck function and~$\tau_\nu$ the dust optical depth (which is proportional to the dust absorption coefficient, $\kappa_\nu\propto\nu^\beta$). In the Rayleigh-Jeans regime and for optically thin emission, we expect $\alpha \approx 2+ \beta \geq 2$. Small ISM-like grains have a~$\beta$ parameter of~1.5-2~\citep[e.g.,][]{Li_Draine_2001}, while grains of millimeter or centimeter sizes are expected to have a~$\beta$ parameter closer to~0~\citep[e.g.,][]{Pavlyuchenkov_2019}.  
Low values of spectral indices~($\alpha \leq 3$) are usually interpreted either in terms of the emission being optically thick or that the dust grain size distribution has grown significantly to reach millimeter/centimeter for the maximum sizes \citep[e.g.,][]{Testi2014}. 

Using the continuum fluxes from our survey and~(sub)millimeter fluxes from the literature, we estimate the global millimeter spectral index~$\alpha_{mm}$ for each source. We use a least-squares fit of all photometric points between~800\,$\mu$m and~3.3\,mm. We find $\alpha_{mm} \leq 3 $ for all disks, as can be seen in~Table~\ref{tab:alpha_glob}.\\

\begin{table}[]
    \centering
    \caption{Integrated spectral indices.}
    \begin{tabular}{cc}
    \hline
    \hline
    Sources & $\alpha_{mm}$\\
    \hline
    \tauZero& $2.5\pm0.1$\\
    \hh& $2.5\pm 0.1$\\
    \IRAS&  $2.3\pm 0.1$\\
    \hk &$3.0\pm 0.2$\\
    \hv &$2.2\pm0.2$ \\
    \JTroisZeroSept& $2.1 \pm 0.4$\\
    \JDeuxDeux&  $2.6 \pm 0.1$ \\
    \irasGerrit& $2.9 \pm 0.6$ \\
    \Oph& $2.6\pm0.1$\\
    \Junun& $2.3\pm0.2$\\
    \hline
    \end{tabular}
    \tablefoot{The millimeter spectral indices were calculated using the fluxes from Table~\ref{tab:results_fluxes} together with literature measurements (references for these are reported below). No spectral indices are reported for \Junsois\: and \hhq\: because they were observed in only one millimeter band.}
    \tablebib{\tauZero: \citet{Andrews_2013},  \hh: \citet{Louvet_2018}, \IRAS: \citet{Moriarty-Schieven_1994, Grafe_2013, Wolf_2003}, \hk: \citet{Duchene_2003}, \hv: \citet{Andrews_2013, Duchene_2010}, \JTroisZeroSept: \citet{Andrews_2013},  \JDeuxDeux: \citet{Dutrey_1996}, \irasGerrit: \citet{Andrews_2008, Motte_2001}, \Oph: \citet{Cox_2017}, \Junun: \citet{Wolff_2017}}
    \label{tab:alpha_glob}
\end{table}

Global spectral indices do not take into account the spatial distribution of the emission. Thus, for sources with multiple millimeter images, we computed spectral index maps using the band~4 and band~7 observations.  
We generated the spectral index maps pixel-by-pixel by applying the CASA task \texttt{immath} on the band~7 and band~4 maps computed to a unique resolution~(see Section~\ref{sec:ALMAimaging}).  
Finally, to lower the noise level in the spectral index map, we applied a filter to keep only the pixels with emission well above the noise level~(5$\sigma$) in both the band~4 and the band~7 images. The final maps and cuts along the major and minor axis are displayed in~Fig.~\ref{fig:alpha_map_cuts}.

We find that the spectral index increases with radius for most sources, albeit with larger uncertainty at larger radii due to the lower signal-to-noise ratio. Similarly, the spectral index also rises along the minor axis direction for two disks, \tauZero~and \JTroisZeroSept.  Previous studies identified similar increases in the radial direction in several disks seen at lower inclinations~\citep[e.g.,][]{Pinte_2016, Dent_2019}. The increases were attributed to changes in the dust size distribution and/or to lower optical depths at large radii. Very inclined systems on the other hand appear optically thicker than low inclination ones for the same mass. This is due to projection effects, because the line-of-sight crosses the disk over a longer distance. This suggests that, for the most inclined disks of our sample, spectral index variations are dominated by changes of optical depth inside the disk (i.e., opacity effects) rather than by grain growth (see also our radiative transfer model, Appendix~\ref{apdx:models}). However, at the outer edges (both radially and vertically), where the disks become optically thinner, spectral index variations are enhanced by changes in the dust size distribution~(see e.g., \tauZero, \IRAS, \JTroisZeroSept, and \JDeuxDeux, where we found large differences in major axis sizes between band~7 and band~4). 
We also note that \hh, \IRAS, \hk, and \hv~show the opposite behavior of \tauZero~and \JTroisZeroSept\:along the minor axis, their spectral indices decrease. However, none of them is well resolved along the minor axis at the resolution of the restoring beam, so variations can be more affected by beam dilution and are less reliable. 

Finally, we point out that we obtain spectral index values lower than 2 in the innermost regions of two disks: namely \tauZero\:and~\JTroisZeroSept~(see also our radiative transfer model, Appendix~\ref{apdx:models}). Such low values have also been reported in the inner regions of other disks~\citep[e.g.,][]{Huang_2018, Dent_2019}, and have often been interpreted as flux calibration errors because in the Rayleigh-Jeans regime~$\alpha$ should not be smaller than~2. 
However, considering a~10\% flux calibration error~(yellow shaded regions in~Fig.~\ref{fig:alpha_map_cuts}), the lowest~$\alpha$ values measured in \tauZero~and~\JTroisZeroSept~can not be reconciled with~$\alpha=2$.
Recent studies showed that low dust temperature (outside the Rayleigh Jeans regime) or dust scattering in optically thick regions can reduce significantly the emission of a disk. In both cases, the spectral index can be even lower than 2~\citep[e.g.,][]{LiuBaobab_2019, Zhu_2019}. In the context of highly inclined disks such as \tauZero~and~\JTroisZeroSept, which are optically thick, both explanations are equally valid to explain low spectral indices observed. Modeling is needed to determine which one is dominant. 

To summarize, because of the high inclination of our sources, we interpret the observed variation of spectral indices along the major axis as being dominated by optical depth changes in the disks.  Additionally, we find that the low spectral index values measured in \tauZero~and~\JTroisZeroSept\:can either be related to low dust temperatures or to dust scattering in optically thick regions. 

\begin{figure*}
    \centering
    \includegraphics[width =18cm]{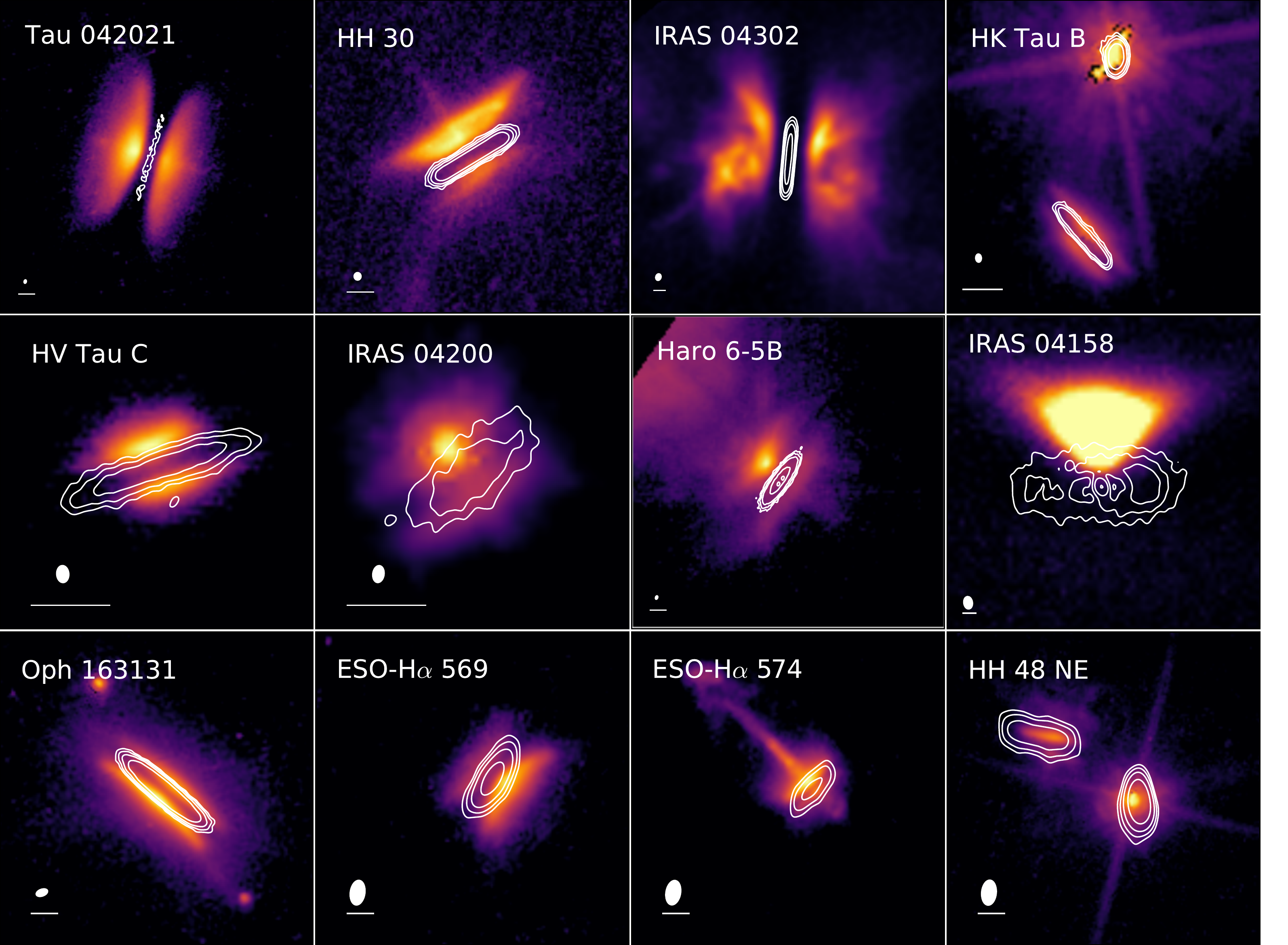}
    \caption{Overlay of scattered light (colors, in logarithmic scales) and ALMA band~7 continuum images (5, 10, 20, and 50$\sigma$ contours) for all sources in this study, except for \Oph~for which we show the band~6 image. The scattered light images are plotted between 3$\sigma$ (except for \IRAS~and \irasGerrit, where we use respectively 60 and 1$\sigma$ to increase the contrast) and their maximum brightness (except for \irasGerrit, for which we take a lower value to increase the contrast). The ellipse and horizontal line in the bottom left corner indicate the beam size of the ALMA image and a 0.5\arcsec scale.}
    \tablebib{Scattered light images, \textbf{2.2\,$\mu$m:} \hv: \citet{Duchene_2010}; \textbf{1.9\,$\mu$m:} \IRAS: \citet{Padgett_1999}; \textbf{1.6\,$\mu$m:} \JDeuxDeux: \citet{Padgett_1999}; \textbf{0.8\,$\mu$m:} \tauZero: \citet{Duchene_2014}, \hh: \citet{Watson_Stapelfeldt_2004}, \irasGerrit: \citet{Glauser_2008}, \Junun: \citet{Wolff_2017}, \hhq: \citet{Stapelfeldt_2014}; \textbf{0.6\,$\mu$m:} \JTroisZeroSept: Stapelfeldt et al. (in prep), \Oph: \citet{Stapelfeldt_2014}, \Junsois: \citet{Stapelfeldt_2014};  \textbf{0.4\,$\mu$m:} \hk: Duch\^{e}ne et al. (in prep).}
    \label{fig:overlay_alma_hst}
\end{figure*}

\subsection{Comparison with optical and NIR images}
\label{sec:HST_compare}
We present overlays of optical and NIR images with our band~7 observations in~Fig.~\ref{fig:overlay_alma_hst}. For most disks, we use HST optical images but prefer (space-based or ground-based) near-infrared~(NIR) images in a handful of cases to reduce confusion with extended nebulosity~(see references in Fig.~\ref{fig:overlay_alma_hst}). 
All scattered light images show the same characteristic features, with two bright reflection nebulae separated by a dark lane tracing the obscuration of direct starlight by the edge-on disk. 
As opposed to the scattered light images, the millimeter data appear as very flat disks. All sources are found much less extended vertically in the millimeter than in scattered light, the result of a combination of opacity effects and vertical settling. Most of them also appear less extended radially, which can be linked to dust radial drift or opacity effects. 

We estimate the scattered light major axis sizes by following the spine of each nebula~(see method in Appendix~\ref{sec:estimation_hst_size}) and report the inferred radial sizes in~Table~\ref{tab:major_axis_cut}. We could not estimate the scattered light sizes for two disks of the sample: for \IRAS\:because we do not see the disk but the envelope; and for \JTroisZeroSept\:because the bright point source in the northern nebulae prevented the method to converge. Also, we indicate that our scattered light radial sizes might be under-estimated because lower illumination or lower sensitivity in the outer regions might reduce the apparent optical-NIR size~\citep[see for example][]{Muro-Arena_2018}.
A complete analysis will require the use of tracers of the gas distribution, which we postpone to a future paper.

Despite this, we find that most sources are larger radially in scattered light than in the millimeter, albeit with a few exceptions, the most obvious case being \hv. \Junsois, and \hhq,
although formally more compact in scattered as estimated with our algorithm, have very similar sizes and will require deeper millimeter data for confirmation.

The ratios of scattered light over thermal continuum band~7 sizes are between 0.7 and~2.0. This is in general consistent with predictions from radial drift theory: the objects with clear evidence for radial drift being those in which scattered light disk~(small grains) is significantly larger than the millimeter continuum disk~(large grains). While we do not quantify explicitly the vertical extent of the scattered light images, Fig.~\ref{fig:overlay_alma_hst} clearly shows that all disks are significantly more extended vertically in scattered light than at millimeter wavelengths, which indicates vertical settling has occurred in each disk.

\section{Discussion}
\label{sec:discussion}

\begin{figure*}
    \centering
    \includegraphics[width =\textwidth]{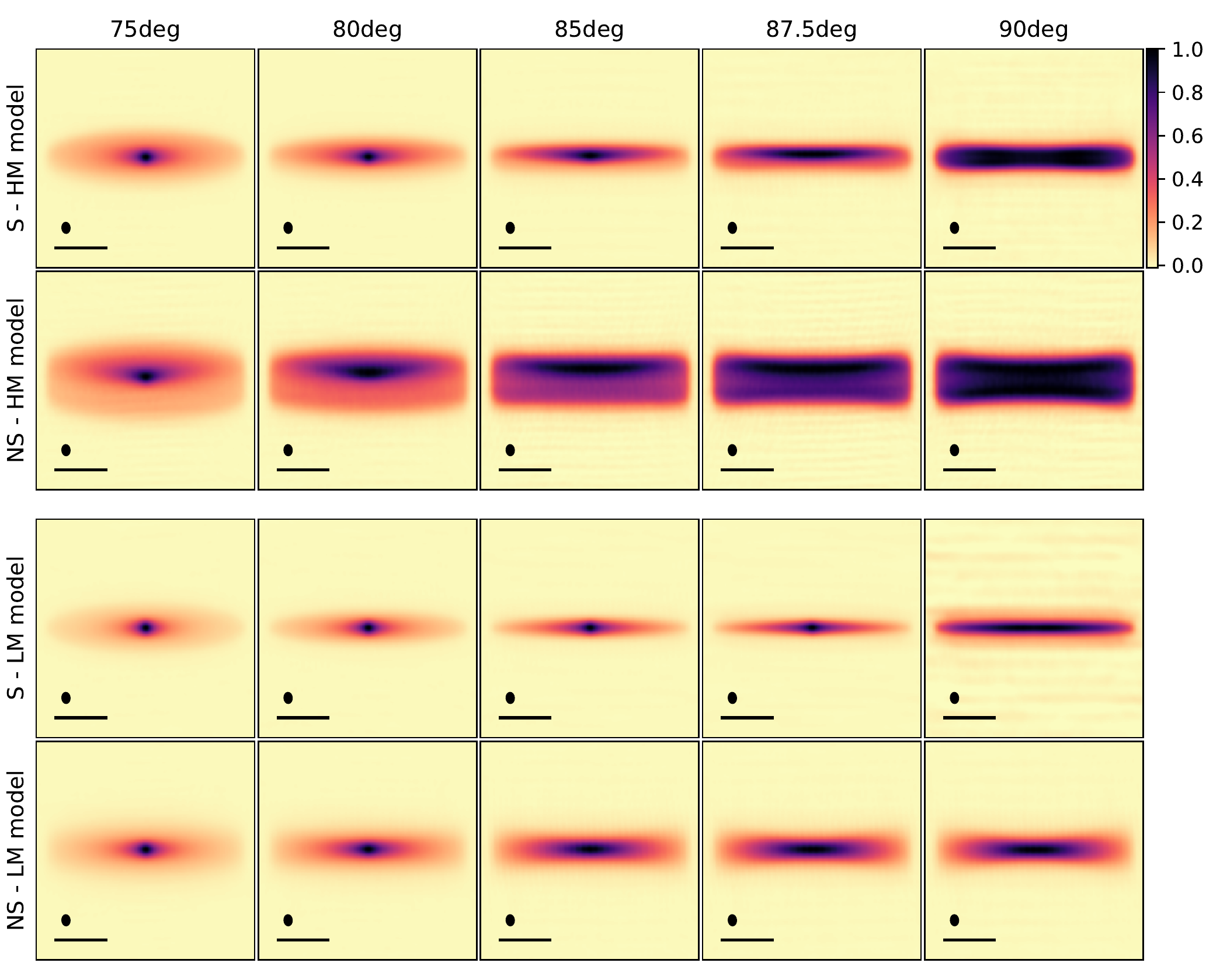}
    \caption{Radiative transfer models computed at 0.89\,mm (band~7), for different inclinations. \emph{Top and third row:} High and low mass settling models (h$_\mathrm{1mm}=0.70$\,au at r=100\,au), \emph{Second and bottom row:} High and low mass No Settling models (h$_\mathrm{1mm}=\mathrm{h}_\mathrm{gas}=10$\,au at r=100\,au). The beam size and a 0.5\arcsec\:scale bar are shown in the bottom left corner of each panel.}
    \label{fig:modelHLTau}
\end{figure*}

\label{sec:model}

In this section, we use the brightness profiles of the highly inclined disks of our sample to discuss critically the amplitude of vertical dust settling, radial drift, and the effects of the enhanced optical depth due to projection effects. To this effect we constructed several toy disk models and performed radiative transfer with the \textsc{mcfost} code~\citep{Pinte_2006, Pinte_2009} to produce synthetic images for comparison with the data. The toy models include a disk without rings or gaps and have an outer radius of 140\,au (1\arcsec). We assume that the surface density follows a truncated power law. The synthetic images are computed with and without vertical dust settling. 

For the model without settling~(NS), the dust is well mixed with the gas and assumed to have a scale height of 10\,au at a radius of 100\,au. For the model with dust settling~(S), we assume that, as a function of size, dust follows the vertical density profile prescribed by \citet{Fromang_Nelson_2009}. Following the results from \citet{Pinte_2016} for HL\,Tau, a very flat disk when observed with ALMA, we set the vertical distribution of the millimeter dust (expressed in terms of a "scale height") to be h$_\mathrm{1mm}=0.7$\,au at 100\,au. This corresponds to a disk viscosity coefficient of~$\alpha = 3\cdot10^{-4}$. 

The toy model is also calculated for two different dust masses in order to probe the effect of optical depth. We consider intermediate to high mass disk models based on the upper limits on the dust mass derived for our sample of highly inclined disks (Section~\ref{sec:flux+derivedMasses}). We use $M_\mathrm{dust}=1\cdot10^{-3}\,M_\odot$ for high mass disk models (HM), and $M_\mathrm{dust}=5\cdot10^{-5}\,M_\odot$ for low mass models (LM).
In total four different sets of images are calculated. We use a distance of 140\,pc.

We computed the models at~0.89\,mm~(resp. 2.06\,mm) for inclinations between~{75}$^\circ$ and~90$^\circ$, and produced synthetic images using the CASA simulator with the same uv-coverage as our band~7 (resp. band~4) data. 
The synthetic band~7 images~(0.89\,mm) of each model are presented in~Fig.~\ref{fig:modelHLTau}.  The band~4 images are qualitatively similar as those presented in Fig.~\ref{fig:modelHLTau} but slightly less extended in the vertical direction.

We also generated band~7 and band~4 model images with a unique angular resolution using a \texttt{uv-taper}. Using these maps, we computed the brightness temperatures and spectral index maps of the models, which are presented in Appendix~\ref{sec:appendix_T} and~\ref{apdx:models}.

\subsection{Constraints from the surface brightness profiles}
\label{sec:verticalExtent}
\subsubsection{Vertical extent}
\label{sec:scaleHeight}

\begin{figure}
    \centering
    \includegraphics[width =0.5\textwidth]{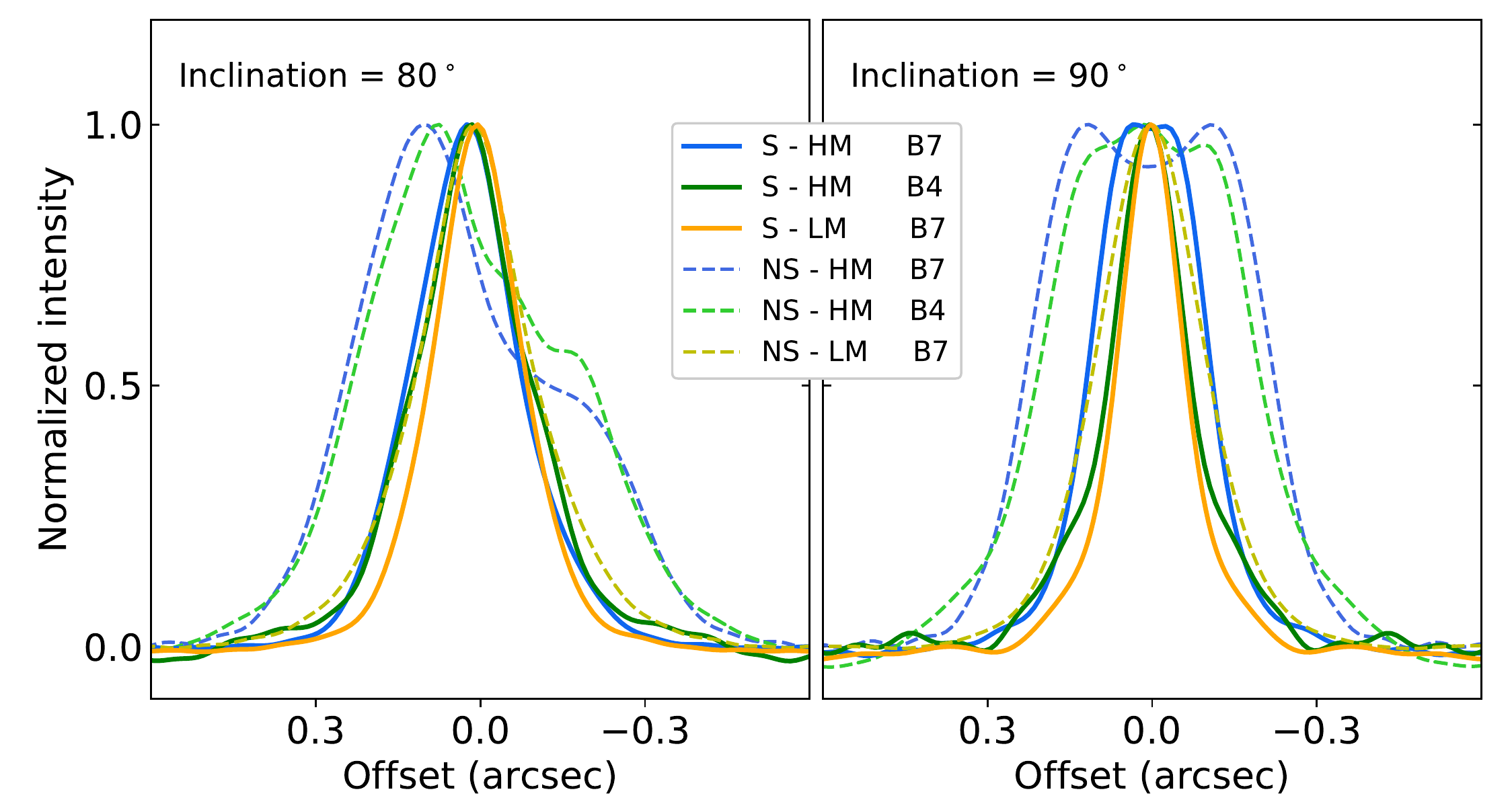}
    \caption{Normalized averaged minor axis profiles for the high and low mass settling models (solid lines) and  No Settling models (dashed lines) computed at 0.89\,mm (band 7), at 80$^\circ$ (left panel) and 90$^\circ$~(right panel).}
    \label{fig:model_minAxis}
\end{figure}

For the high and low mass models presented in  Fig.~\ref{fig:modelHLTau}, we see that the disks appear less extended in the minor axis direction when settling is included. For the high mass model without settling~(model NS-HM), we find that as the inclination approaches~90$^\circ$, the high optical depth in the midplane results in a clear low intensity lane, separating the two sides of the disk. On the other hand, at lower inclinations the bottom~(far)~side of the disk is about~5-6~times fainter than the top~(near)~side. This is visible directly in Fig.~\ref{fig:modelHLTau} and highlighted in a different way by showing cuts along the minor axis in Fig.~\ref{fig:model_minAxis} for inclinations of 80$^\circ$~(fainter back side) and 90$^\circ$ (dark lane). As the angular resolution is similar and the signal-to-noise is greater than~14 for all targets included in this study, such an asymmetry would be detectable easily in our observations. However, neither the vertical asymmetry nor the dark lane in the midplane are present in our data. This configuration (NS-HM) can be ruled out. Interestingly, the features can hardly be seen in the high mass model which includes settling, nor in any of the low mass models. If the disks included in our sample are as massive as the high mass model, this would indicate that they have a small millimeter dust scale height, closer to 1\,au than to 10\,au at a radius of 100\,au. This is similar to the results of \citet{Pinte_2016} for the disk of HL\,Tau.

To go further, the deconvolved minor axis sizes of the models can be compared with the average size obtained for the most inclined disks of our sample (i.e., with $i> 80^\circ$).  For the data, the mean deconvolved minor axis size is about 0.18\arcsec\:in band~7 (0.17\arcsec\:in band~4, see Table~\ref{tab:radial_cuts_min}). Except for the high mass model without settling (model NS-HM), which is more than twice as thick as the observations~(deconvolved minor axis size $\sim 0.4\arcsec$ at 80$^\circ$ and 90$^\circ$), all models studied in this section are compatible with the vertical extent measured in the data~(in band~7, typical deconvolved sizes of $\sim 0.2\arcsec$ for S-HM, NS-LM at 80$^\circ$ and 90$^\circ$, and for S-LM at 80$^\circ$, and $\sim 0.1\arcsec$ for S-LM at 90$^\circ$). Thus, the vertical extent alone is not sufficient to distinguish whether vertical settling is required to explain the observations or not.

\subsubsection{Radial brightness profile}
\label{sec:radialHLTau}
\begin{figure}
    \centering
    \includegraphics[width =0.5\textwidth]{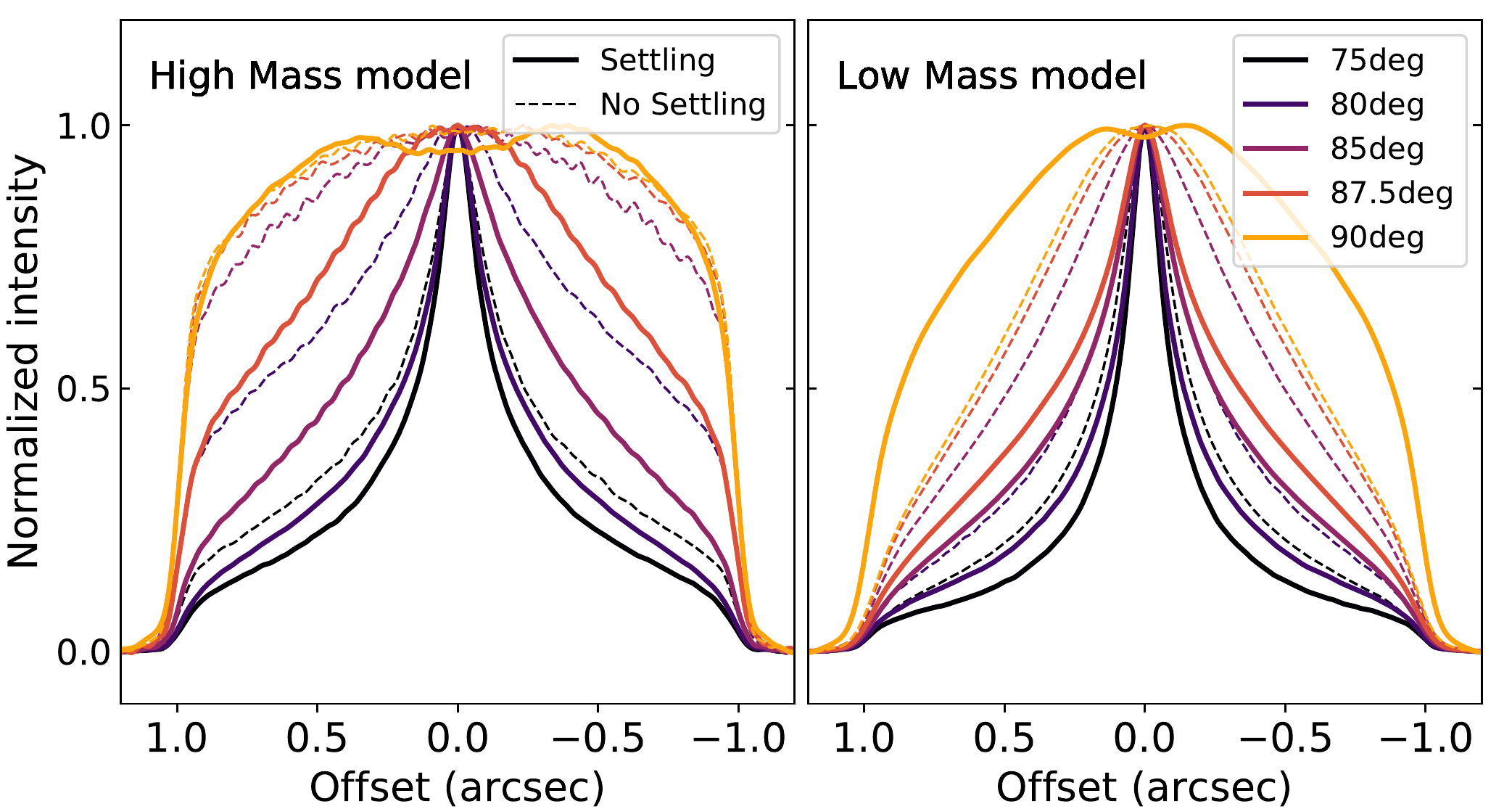}
    \caption{Normalized major axis intensity profiles for the high mass (left) and low mass (right) models at 0.89\,mm, for different inclinations between 75$^\circ$ and 90$^\circ$.}
    \label{fig:model_majAxis}
\end{figure}

We now investigate the effect of inclination on the observed radial brightness profiles.
To do so, we produce major axis cuts of our radiative transfer models, as in Section~\ref{sec:radial_extent}, and present them in Fig.~\ref{fig:model_majAxis}. We also estimate the apparent sharpness between 20\% and 80\% of the peak flux ($\Delta r/r$) for our models as in Section~\ref{sec:radial_extent}, and report them in~Table~\ref{tab:models}.

Along the major axis, the effect of inclination on the shape of the brightness profile is very clear both in the images~(Fig.~\ref{fig:modelHLTau}) and in the cuts (Fig.~\ref{fig:model_majAxis}). For all models, at the lowest inclinations, the images and cuts show a steep increase in intensity at the center, and low apparent edge sharpness ($\Delta r/r >1$ for $i<80^\circ$). 
On the contrary, for the fully edge-on configuration~($i=90^\circ$), the major axis brightness profile of the high mass models is flat at all radii and drops steeply ($\Delta r/r \sim 0.2$ at $90^\circ$ for both high mass models S-HM and NS-HM). 
Between these extreme cases, the cuts show a less extended plateau and shallower outer edges than for the 90$^\circ$ case, related to lower optical depth than in the edge-on case. 
We note that in the low mass models, which are optically thinner, the flat plateau along the major axis direction is never reached and the apparent disk sharpness is always larger than $\Delta r/r  >0.6$.  
Nevertheless, we find that, independently of the dust mass assumed for the models, the apparent sharpness of the disk outer edge increases with increasing inclination.\\
 
Three disks in our sample show edges as sharp as~0.3. They are \hh, \hk, and \hhq. The comparison with models suggests that these disks are the most inclined of the sample.  Additionally, these disks present flat radial brightness profiles that the low mass models are unable to reproduce. This indicates that they are optically thicker than the low mass model, likely because they are more massive. In Section~\ref{sec:scaleHeight}, we have shown that a more massive disk, while leading to a flat profile along the major axis direction also presents a larger vertical extent. Our high mass model without settling is inconsistent with the observations (see Fig.~\ref{fig:modelHLTau}, NS-HM model). This implies that vertical settling is needed to explain both the flat radial profile and small apparent vertical extent of these three disks. For these systems the millimeter scale height would be closer to 1\,au than to 10\,au at 100\,au, therefore increasing significantly the concentration of dust mass in the disk midplane.
Although we can not confirm vertical settling for the other sources, we believe that their vertical structure is likely similar and governed by settling, in particular because of the large vertical size difference between the scattered light and thermal emission images.

Vertical settling models show that the turbulence generated self-consistently by ideal MHD or vertical shear instabilities are likely too strong to lead to millimeter scale height as small as 1\,au at r = 100\,au for gas scale heights of 10\,au at 100\,au~\citep{Flock_2017, Flock_2020}. On the other hand, non-ideal effects such as ohmic resistivity or ambipolar diffusion lead to lower turbulence and thus to very thin millimeter grain layers~\citep{Riols_2018}. Those mechanisms may well be dominant in the sample of disks analyzed here. Detailed modeling of each individual object is needed to obtain a quantitative estimate of the millimeter dust and gas scale height of the disks.

\begin{table}
\caption{Effect of inclination on the apparent edge sharpness $\Delta r/r$ for our radiative transfer models.}
\centering
\begin{tabular}{lcccc}
\hline
\hline
Incl (deg)& S - HM & NS - HM & S - LM & NS - LM\\
\hline
75 & $1.7\pm0.3$ & $1.7\pm0.2$ & $1.5 \pm 0.4$ & $1.5 \pm 0.3$\\
80 & $1.7\pm0.2$ & $1.2\pm0.1$& $1.6\pm 0.3$ & $1.6 \pm 0.3$\\
85 & $1.5\pm0.1$ & $0.4\pm0.1$ & $1.6\pm 0.2$ & $1.2 \pm 0.1$\\
87.5 & $0.9\pm0.1$ & $0.2\pm0.1$ & $1.6 \pm 0.2$ & $1.0 \pm 0.1$\\
90 & $0.2\pm0.1$ & $0.2\pm0.1$ & $0.6 \pm 0.1$ & $1.0 \pm 0.1$\\
\hline
\end{tabular}
\label{tab:models}
\end{table}

\subsection{Comparison with a radial drift model}
Similarly to previous multi-wavelength studies~\citep[e.g.,][]{perez_2012, Tripathi_2018, Powell_2019}, our observations show that 4 disks have major axis sizes that decrease with wavelength (namely \tauZero, \IRAS, \JTroisZeroSept, \JDeuxDeux, see Table~\ref{tab:major_axis_cut} and the radial variation of their spectral index in Fig.~\ref{fig:alpha_map_cuts}). For these disks, the average size difference between band~7 and band~4 observations is about 12\%. Estimating the major axis sizes from our radiative transfer models described in Section~\ref{sec:model}~(which do not include radial drift), we find that opacity effects alone predict a difference of only a few percents between bands, which is not sufficient to reproduce the observations.  
In this section, we compare the measured radial differences with an analytical radial drift model, presented by \citet{Birnstiel_2014}. Similarly to other theoretical models, they showed that inward dust migration of single size particles spontaneously produces a sharp edge in the dust density distribution~\citep{Birnstiel_2014, Facchini_2017, Powell_2019}. \citet{Birnstiel_2014} computed an analytical formula to infer the position of the disk outer edge in a disk with radial drift only (see their equation B9). The vertical extent of the grains is not considered in their model.  
They assume a smooth tapered-edge gas surface density profile and parametrize the turbulence following~\citet{Shakura_Sunyaev_1973}.

Assuming that the band~7 and band~4 emission only originate from grains of the optimal size~($a\approx\lambda/2\pi$), the model by \citet{Birnstiel_2014} predicts a size difference in surface density between band~7 and band~4 of about~25\%~(after 0.1\,Myr). 
While the predicted effect is marginally too strong, the absolute disk sizes are more problematic. When grains of~0.1\,$\mu$m detectable in scattered light are expected to be found up to~135\,au after~0.1\,Myr, the model predicts that grains emitting most at~0.89\,mm should have drifted to~42\,au. This corresponds to a micron/millimeter disk radius ratio greater than~3, which is more than 1.5 times larger than the largest ratio measured in our data. 
Part of these differences might be explained because the observations are not probing the surface density of the disk, because several grain sizes (that might have drifted to different radii) have to be considered rather than a unique one, or because our disks are older than 0.1\,Myr so they might be affected by viscous spreading as well. 
Besides, as discussed previously~(Section~\ref{sec:HST_compare}), we note the scattered light images might not always trace the whole disks, since they require illumination (and sufficient optical depth) to trace the disk all the way to the edge. This can lead to apparent sizes in scattered light that are smaller than the real radial extent of small grains. 
However, the small expected sizes at millimeter wavelengths suggest that this radial drift model is too efficient to reproduce the observations~\citep[see also][]{Brauer_2007}. The existence of pressure fluctuations in the disks (e.g., rings and gaps tracing pressure bumps and dust traps) rather than smooth power law surface density profiles are expected to slow down the drift efficiently~\citep[see, e.g.,][]{Gonzalez_2017, Pinilla_2012} and help reconcile models with observations. We speculate that the disks included in our study may have complex radial structures to slow down the radial migration of dust.

\subsection{Effect of inclination on global values}
\label{sec:integratedValues}

Most Class~II disk surveys in close-by star-forming regions estimate dust masses directly from the measured integrated fluxes assuming optically thin disks, irrespective of the disk inclinations~\citep[e.g.,][]{Andrews_2013, Ansdell_2016, van_der_Plas_2016}. However, because of the projection effect, the optical depth along the line-of-sight increases with inclination and will affect the observed flux. In this section, we investigate the impact of inclination on integrated fluxes, dust mass estimations, and integrated spectral indices. Finally, we also discuss the measured brightness temperatures obtained for our disks in the context of optical depth.

\subsubsection{Flux density and derived masses}
\label{sec:flux+derivedMasses}
\begin{figure}
    \centering
    \includegraphics[width =9cm]{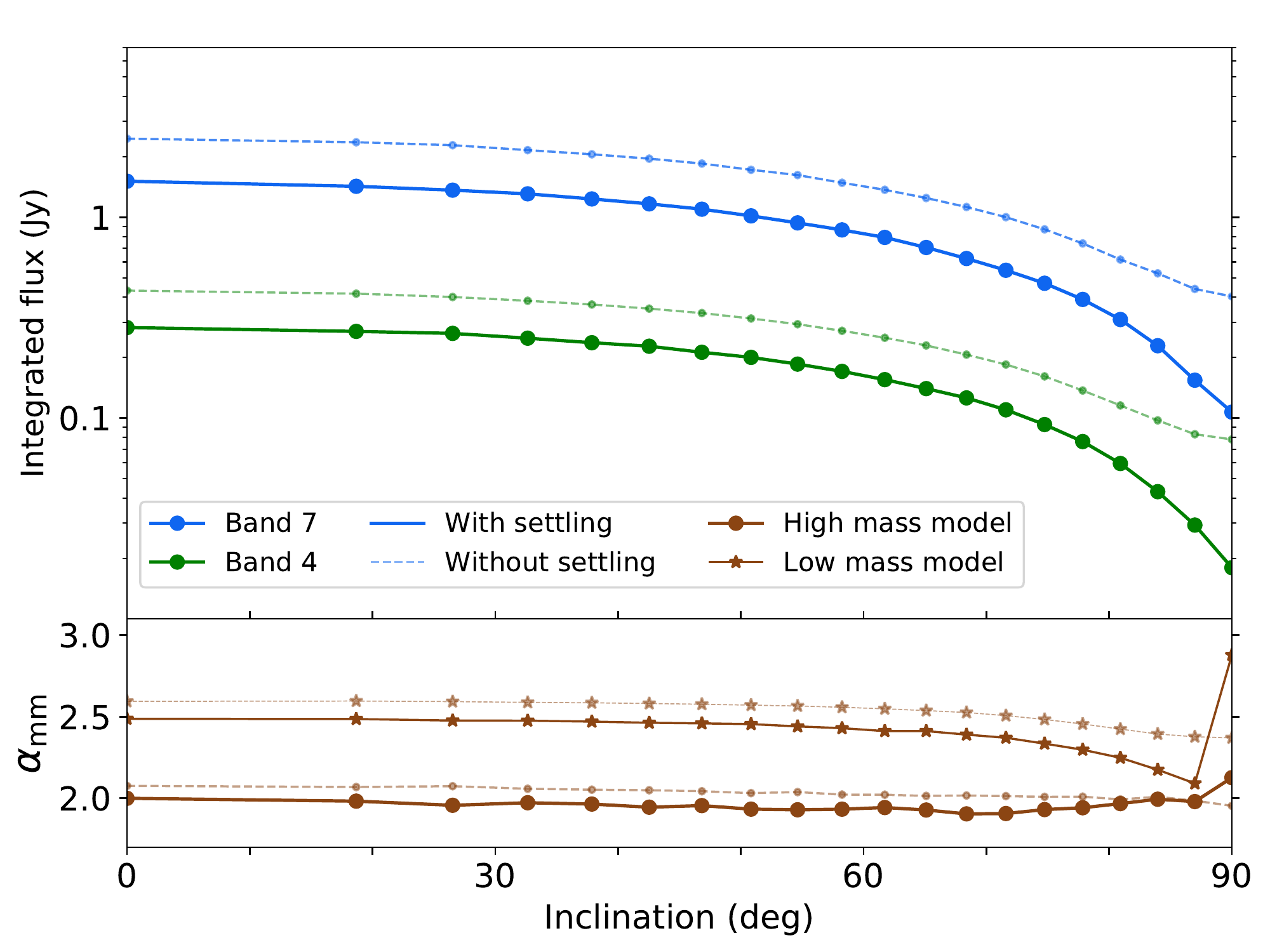}
    \caption{\emph{Top:} Variation of the band~7 and band~4 integrated fluxes of the high mass radiative transfer models with settling (thick lines) and without settling (thin lines) as a function of the inclination. \emph{Bottom:} Variation of the integrated spectral index with inclination for both high and low mass radiative transfer models, with or without including settling.}
    \label{fig:flux_incl}
\end{figure}

We present the variation of the integrated band~7 and band~4 fluxes in our high mass radiative transfer models~(with and without settling) as a function of inclination in~Fig.~\ref{fig:flux_incl}.  The effects of dust scattering are fully included in the radiative transfer calculations. 
Overall, the emitted flux density of the disk becomes attenuated by up to an order of magnitude with increasing inclination. This is due to a combination of the increasing optical depth, a lower average dust temperature seen by the observer for high inclinations, and to geometrical effects~(reduced emitting surface with increasing inclination). 
In this model, from  $\sim$60$^\circ$ to 90$^\circ$, the flux density and thus the derived disk mass would appear significantly smaller than for the same disk viewed at intermediate inclinations~(by up to $\sim$10 times for the high mass settled model). The amplitude of the attenuation depends on the parameters of the model. 
We note that in Fig.~\ref{fig:flux_incl}, the model without settling is brighter than the corresponding settled model for all inclinations. This is due to temperature differences between the settled and no settling models, to differences in optical depths and to geometrical effects. Similar behaviors are seen in the low mass models which are not represented. \\

\begin{table}[]
    \centering
    \caption{Mass limits.}
    \begin{tabular}{ccccc}
    \hline
    \hline
    Sources & M$_\mathrm{dust\,B7}$ (M$_\oplus$)\\
    \hline
    \tauZero&  > 25.5\\
    \hh&  > 11.2\\
    \IRAS&  > 54.8\\
    \hk & > 11.4\\
    \hv& > 18.6\\
    \JTroisZeroSept& > 13.4\\
    \JDeuxDeux&   > 69.8\\
    \irasGerrit & > 53.1\\
    \Oph&  > 25.8\\
    \Junun& > 13.6\\
    \Junsois& > 3.1~\\
    \hhq& > 7.1~\\
    \hline
    \end{tabular}
    \tablefoot{Dust masses are estimated from the band~7 fluxes~(Table~\ref{tab:results_fluxes}), assuming optically thin emission, and therefore are lower limits (see text for details).}
    \label{tab:dust_masses}
\end{table}

This raises the question of the reliability of disk masses derived purely from millimeter fluxes, in particular for surveys where the disks are not well resolved and inclinations cannot be estimated. We focus here on the highly inclined disks of our sample. For a direct comparison with previous studies, we assume optically thin and isothermal dust emission at sub-millimeter wavelengths. In that case the flux ($F_\nu$) is directly related to the dust mass following: $M_\mathrm{dust}=F_\nu d^2 / \kappa_\nu B_\nu(T_\mathrm{dust})$, where $d$ is the disk distance to the Sun, $B_\nu(T_\mathrm{dust})$ the Planck function evaluated at a mean representative dust temperature $T_\mathrm{dust}$, and $\kappa_\nu$ the \emph{dust} grain opacity. We use typical values of~$T_\mathrm{dust}$ = 20\,K and~$\kappa_\nu=3.4$\,cm$^2\,$g$^{-1}$, as in \citet[][and references therein]{Ansdell_2016} to estimate the dust mass of the disks in the sample, using the band~7 fluxes. For the data, the results are reported in Table~\ref{tab:dust_masses}. 
As noted above, these masses are almost certainly underestimated and so we quote them as lower limits. 

By applying the formula directly we find a mean dust mass of about~25\,M$_\oplus$ for the~6~edge-on disks more inclined than~75$^\circ$. For comparison, at an inclination of 90$^\circ$, the estimated dust mass of both the high mass and low mass settled models are more than 3 times lower than the real dust mass. 
Without full modeling, it is difficult to find a reliable correction factor for individual sources to compensate the attenuation due to inclination. The disk masses of our highly inclined sample are probably a few times larger, that is up to~$\sim$75\,M$_{\oplus}$ when applying the same factor of~3. This is on the high end of the dust mass distributions of Taurus or Lupus star-forming regions~\citep[mean of~15\,M$_\oplus$,][]{Ansdell_2016}. 

This result suggests that our sample of edge-on disks is biased toward more massive disks. This is a direct consequence of the important attenuation caused by the high inclination, making the starlight and the disk emission fainter.   This may also explain, at least partly, why the number of known edge-on disks with resolved images remains sparse, even today. A comprehensive study of the biases affecting the edge-on disk population and their detection will be presented in Angelo et al.~(in prep).

\subsubsection{Integrated spectral indices}
\label{sec:int_alpha}

The median spectral index of the edge-on disks in our study, $\bar{\alpha}_{mm}=2.5\pm0.3$, is similar to previous measurements of intermediate-inclination disks, for example $2.3 \pm 0.1$ in the Lupus and Taurus star-forming regions \citep{Ribas_2017, Ansdell_2018}. At first glance, this is surprising as edge-on disks are expected to appear more optically thick, but \citet{Sierra_2020} showed that the spectral index is only mildly dependent on inclination based on an extensive study.

To further understand the integrated $\alpha_{mm}$ for edge-on disks, we computed them for our radiative transfer models. The expected values at different inclinations are shown in the bottom panel of~Fig.~\ref{fig:flux_incl}.  We find that the variation of spectral index with inclination is relatively small~($1.9 \lesssim \alpha_{mm} \lesssim 2.2$ for the high mass settled model, and $2.1 \lesssim \alpha_{mm} \lesssim 2.5$ for the low mass settling model for inclinations lower than 87$^\circ$), in agreement with \citet{Sierra_2020}. This is because the models are (at least partially) optically thick even at low inclination and because large grains are included (maximum grain size is 3\,mm). 
The small difference in integrated spectral index observed between our sample of edge-on disks and disks at lower inclinations can be explained if most disks are partially optically thick at millimeter wavelengths and/or if grains have grown to millimeter/centimeter sizes. Grain growth is known to have occurred in class~II disks, leading to low spectral indices~\citep{Ricci2010, Testi2014}. Similarly, recent imaging campaigns have revealed that ring structures are very common in protoplanetary disks~\citep{Huang_2018_annular} and generally associated with optically thick regions with large grains~\citep[e.g.,][]{Dent_2019}. So far, the edge-on disks in our sample appear similar to the disks observed on other surveys, except from their viewing angle. 

Interestingly, for inclinations greater than 70$^\circ$ for the high mass or 87$^\circ$ for the low mass model, we find a clear difference in the integrated spectral index between the models with and without settling. While at these inclinations the spectral index decreases in the models without settling \citep[see also][]{Galvan-Madrid_2018}, we find that the integrated spectral index increases for the settled models, reaching about~2.2 at~90$^\circ$ for the high mass model~(2.8 for the low mass model). This can be explained by the contribution of the optically thick midplane that decreases for increasing inclination (as it appears colder) while the contribution of the upper layers (rich in small grains with higher $\beta$ values) increases, leading to higher values of the integrated spectral indices for the settled model~(see, e.g., Fig.~\ref{fig:alpha_model}). Said differently, for very large inclinations several line-of-sights do not cross the disk midplane containing large grains but only the high altitude layers above and below it. Since these layers are optically thinner and contain only small grains, the spectral index increases. However, further studies are required to investigate this effect~\citep[e.g.,][]{Sierra_2020} and its applicability to specific objects. 

\subsubsection{Brightness temperatures}
\label{sec:brightness_T}

Assuming that scattering is negligible and that the dust temperature is high enough (e.g., in the Rayleigh-Jeans limit) and constant over the emitting region, the brightness temperature of dust emission at frequency $\nu$ can be expressed as~$T_B=\eta_c~T_p~(1-e^{-\tau_\nu})$, where~$\tau_\nu$ is the optical depth of the medium, $T_p$~the mean dust physical temperature~\citep[see e.g.,][]{Wilson_2009}, and $\eta_c$ the fraction of beam covered by the source. In the isothermal optically thick limit~($\tau_\nu\gg1$), for a source filling the beam~(but smaller than the largest angular scale of the interferometric observations) the brightness temperature corresponds to $T_p - 2.7$\,K, because the cosmic microwave background~(CMB) is resolved out by the interferometer.
For compact sources, beam dilution would reduce the observed brightness temperatures below~$T_p$. Scattering is also known to decrease dust emission from very optically thick regions, which would also effectively lead to lower observed brightness temperatures~\citep{Zhu_2019}. The brightness temperatures were estimated in Section~\ref{sec:images}. Major axis cuts are presented in~Fig.~\ref{fig:brightness_T} and we report the peak brightness temperatures in Table~\ref{tab:brightness_T}.

We estimate a mean peak brightness temperature of 10.6\,K for the three best resolved sources in band~7 (\tauZero, \JTroisZeroSept, and \JDeuxDeux). These values are particularly low compared to previous estimates on other disks around stars of similar spectral types. 
As an example, \citet{Andrews_2018} derived a mean brightness temperature peak of~66.5\,K for the~17 disks of the DSHARP sample around K\,\&\,M stars observed in band~6. These disks have a mean inclination of~42$^\circ$, while all disks of our study are more inclined than~65$^\circ$. 
The brightness temperatures we derive for our disks are also much lower than the traditional $T_{dust}$ = 20\,K assumed in flux-to-mass conversions~(see e.g., Section~\ref{sec:flux+derivedMasses}).

Although the measured brightness temperatures are integrated over some vertical extent because of the beam size, and therefore include a vertical temperature gradient, for the well-resolved disks $T_B$ does provide a reasonable estimate of the temperature of the outer midplane where the line-of-sight optical depth reaches unity. This is the case for \tauZero\:for example. The low brightness temperatures measured in these optically-thick edge-on systems is likely to reflect the midplane temperature in the cold outer radii of the disks. This in agreement with the idea that the disks appear optically thick because of projection effect.

\subsection{Individual targets}
\subsubsection{\hk: Comparison with a published model}
\label{sec:hk}

\begin{figure*}
    \centering
    \includegraphics[width =\textwidth]{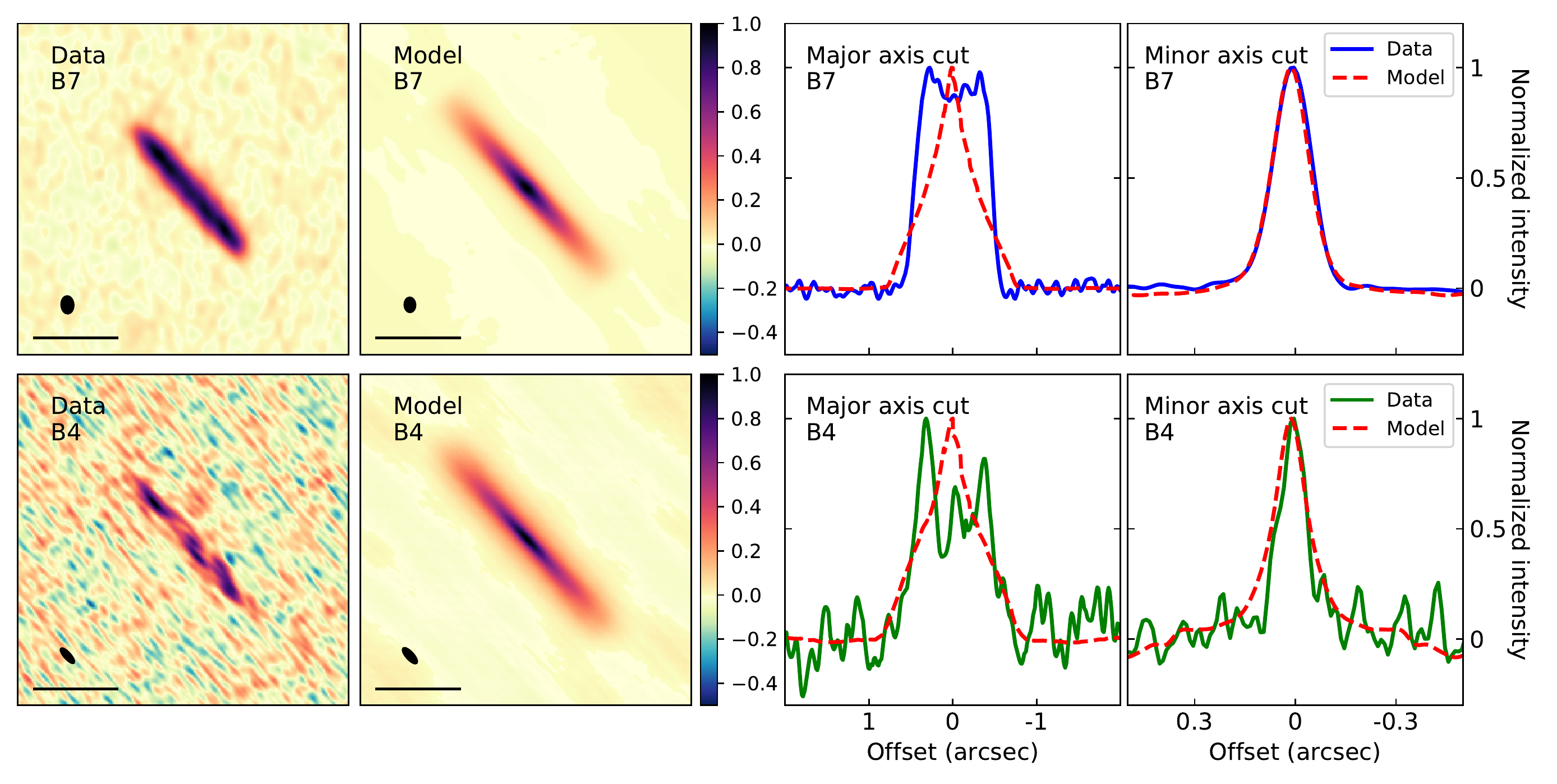}
    \caption{\emph{Left panels:} ALMA data of \hk. In all the figure, top row is band~7 and bottom row is band~4. \emph{Middle-left panels:} Millimeter images of the best model C of \citet{Stapelfeldt_1998}, assuming well mixed grains. They were computed for the same uv-coverage as in the data. \emph{Middle-right panels}: Major axis cuts of the model and the data. \emph{Right panels}: Averaged minor axis cuts of the data and the models, performed as in Section~\ref{sec:vertical_extent}. }
    \label{fig:HKTauB}
\end{figure*}

\hk~is the only Class~II disk of the sample which is resolved by ALMA in the minor axis direction (in band~4) and for which the scattered light image has been modeled previously.  \citet{Stapelfeldt_1998} estimated the scale height of micron-sized grains to be on the order of 3.9\,au at 50\,au~\citep[8.3\,au at 100\,au, assuming a flaring exponent of 1.1 as in][]{Stapelfeldt_1998}. Aiming for a quantitative comparison of scale height between millimeter and micron-sized grains, we computed their best model~C at millimeter wavelengths~(at 0.89\,mm and 2.06\,mm) using the radiative transfer code {\sc mcfost}. The parameters of the model are reported in Table~1 of \citet{Stapelfeldt_1998} and the 0.8\,$\mu$m image of the model can be found in their Fig.~3. 
Following the results of \citet{Duchene_2003}, we adopt revised opacities in our model, leading to a revised disk dust mass of 5.9$\times$10$^{-5}$\,M$_\odot$ to match the observed millimeter fluxes. We use silicate dust grains with sizes from 0.01\,$\mu$m to 15\,cm and assume a number density described with a power law of the grain size $dn(a)\propto a^{-3.5} da$. This yields opacities of 4.8\,cm$^2$/g and 2.6\,cm$^2$/g (per dust mass) respectively at 0.97\,mm and 2.06\,mm. Assuming a gas-to-dust mass ratio of 100, this is comparable to standard assumptions \citep[e.g.,][0.048\,cm$^2$/g and 0.026\,cm$^2$/g of gas and dust, respectively]{Beckwith_1990}.
For a direct comparison, we assume that grains of all sizes are fully mixed and produce synthetic images using the CASA simulator with the same uv coverage as in our observations. We show the results in Fig.~\ref{fig:HKTauB}.

We find that the model does not reproduce well the millimeter images of \hk. Along the major axis, the model is too peaked at the center~(less flat-topped) and more extended in the radial direction than the actual millimeter data. 
As discussed in Section~\ref{sec:radialHLTau}, the more peaked profile suggests that the observed millimeter disk is more optically thick than what is predicted by the model. While the best model of \citet{Stapelfeldt_1998} contains a full disk at an inclination of 85$^\circ$, we suggest that a higher inclination, closer to 90$^\circ$ (see Section~\ref{sec:radialHLTau}), or the presence of a large but unresolved inner cavity would reproduce better the major axis brightness profile.

Along the minor axis, the band~4 cut indicates that the model is also more extended vertically than the data (see Fig.~\ref{fig:HKTauB}).  
As previously proposed by~\citet{Duchene_2003} this hints for vertical segregation of dust grains in this disk, millimeter grains being located in a vertically thinner layer than micron-sized grains. 
However the low signal-to-noise of the band~4 data and the low angular resolution in band~7 prevent us from a strong conclusion.

\subsubsection{Vertical settling in \tauZero}
\label{sec:Tau042021}
\tauZero~is the only edge-on disk clearly resolved vertically in both band~7 and band~4. For this disk, the band~7 appears about~1.5~times more extended vertically than the band~4~(see Table~\ref{tab:radial_cuts_min}). We note that \tauZero\:is only marginally resolved in band~6, so the size at this wavelength is uncertain. 
For our high mass radiative transfer model without settling (i.e., including only optical depth effects), we find a band~7 to band~4 minor axis ratio of about~1.1 at $i=90^\circ$~(respectively 1.2 for the low mass settled model), significantly smaller than the observed value. 
Although this model is not unique, it indicates that opacity effects are not sufficient to produce the large difference in minor axis size observed, which suggests that grain-size-dependent vertical settling is occurring in this disk as well. In this section, we compare the measured minor axis ratio with predictions from several vertical settling models. We assume that the disk is perfectly edge-on so that variations along the minor axis are dominated by differences in vertical extent between bands rather than projections of the disk radius.

From the minor axis ratio measured in \tauZero, one can estimate the scaling of the minor axis size~($S_d$) with grain size~($a$) assuming~$S_d\propto a^{-m}$. Assuming that most of the emission comes from grains of the optimal size~($a\approx \lambda/2\pi$), we obtain: $m = -\log(S_{d, b7}/S_{d, b4}) / \log(a_{b7} / a_{b4}) \sim 0.5$.

If we additionally assume that $S_d$ is directly proportional to the dust "scale height"~(h$_\mathrm{d}$), an exponent of $m=0.5$ has been predicted for large grains in the context of a 1-D diffusion theory~\citep[][]{Dubrulle_1995}. 
Performing numerical simulations including non-ideal MHD effects such as ambipolar diffusion, \citet{Riols_2018} also estimated a relationship of dust scale height with grain size with an exponent of~0.5, valid for large grains~($St \gtrsim 10^{-2}$). For comparison, using the standard disk model presented in \citet[][equation~13]{Riols_2018}, at 100\,au, grains of the optimal size emitting in band~7 would have Stokes numbers larger than~$2.5\cdot10^{-2}$.

On the other hand, \citet{Fromang_Nelson_2009} estimated that in the case of ideal MHD, the dust scale height varies as:~h$_\mathrm{d} \propto a^{-0.2}$. Settling obtained with ideal MHD is expected to be less efficient than other models previously discussed. 
However, we note that this expression was estimated for small grains~($St\leq10^{-2}$) and might not apply for the grain sizes that we are probing at millimeter wavelengths.

To summarize, we find that the large minor axis difference observed in \tauZero\:at different millimeter wavelengths has to be associated with strong settling. To compare our observations with settling models, we make the assumption that the measured minor axis extent is proportional to the real dust scale height. While ideal MHD simulations predict a settling less efficient than observed, a simple 1-D diffusion theory~\citep[in the case of strong settling,][]{Dubrulle_1995} or numerical simulations including non-ideal MHD effects~\citep{Riols_2018} provide for now the best consistency with the observations.

\subsubsection{Radial variation of vertical extent in \IRAS}

\begin{figure}
    \centering
    \includegraphics[width =9cm]{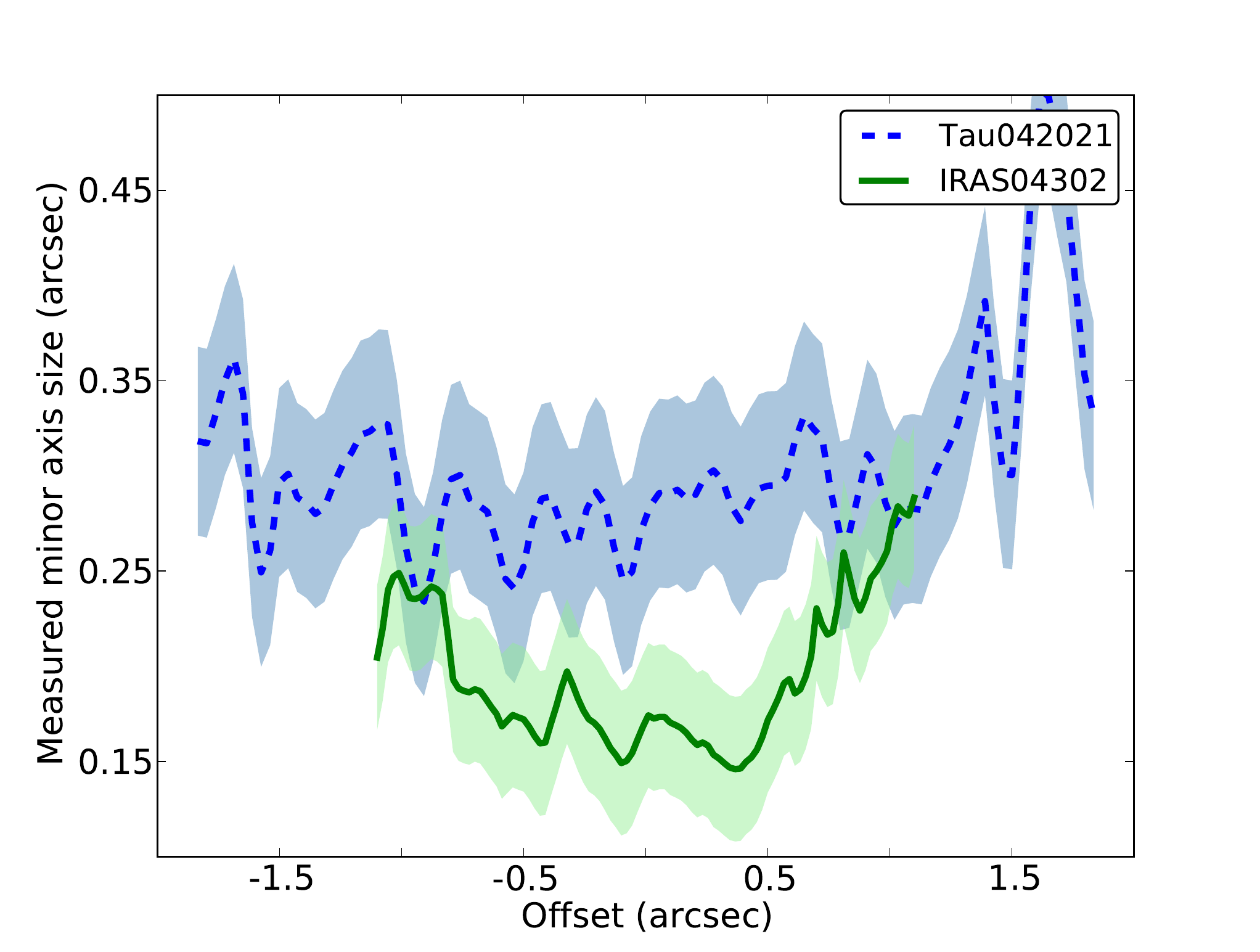}
    \caption{Minor axis size as a function of radius for \IRAS~(band~4) and \tauZero~(band~7). The errors correspond to the standard deviation of the curves.}
    \label{fig:radialVariations}
\end{figure}

The high resolution band~4 disk image of \IRAS\:shows evidence of flaring: the extent perpendicular to the disk midplane at large radii is broader than at the center~(see Fig.~\ref{fig:alma_images}). This is in contrast with the other disks in our survey, which are flatter or unresolved.
Fig.~\ref{fig:radialVariations} compares the minor axis sizes at different radii in \IRAS~with the other well-resolved edge-on disk with large signal-to-noise: \tauZero. In the latter case,  the vertical extent appears quasi constant as a function of radius with only a slight increase in the outer regions, whereas it is $\sim$50\% larger in \IRAS~at a projected radius of 100\,au~(0.7\arcsec) than at the center. 
This might be related to differences in optical depth or settling between the disks. Additionally, \IRAS\:is the only Class~I object in our sample~\citep{Grafe_2013} and the scattered light image does not show the flared upper surface of a disk, like in the other objects~(Class\,II, see~Fig.~\ref{fig:overlay_alma_hst}), but instead traces an envelope.  
Objects like this one, and HH\,212 a Class~0~\citep{Lee_2017}, open the possibility to do comparative studies of disk evolution in the early phases.

\section{Summary and conclusions}
\label{sec:conclusion}

We presented high angular resolution ALMA band~7 and band~4 observations of 12 highly-inclined disks previously selected from the shape of  their scattered light images. All disks are well resolved along the major axis and 6 are also resolved in the direction perpendicular to the disk midplane in at least one millimeter band. Several disks show flat surface brightness profiles along their major axis with a steep drop off at their outer edge, indicating their large inclination and significant optical depth. 

\JDeuxDeux~and \irasGerrit\:are the least inclined disks of the sample~(less than 75$^\circ$) and both show a distinct ring and an isolated emission peak at the center. At the highest angular resolution, the point source at the center of \irasGerrit\:is a binary source. 

The analysis of global quantities such as integrated fluxes, spectral indices, and brightness temperatures shows that the highly inclined disks of our sample have~(at least partly) optically thick emission. Because of the low brightness temperatures and small beam sizes, we conclude that the emission originates from the outer radii of the disks, with a peak brightness temperature below~10\,K for half the sources in band~7.

We also found that the median spectral index in our disk sample is similar to that of disks seen at lower inclinations. This can be explained if disks at intermediate inclination are already partly optically thick~(which implies significant scattering even at millimeter wavelengths) and/or if grains have grown to millimeter/centimeter sizes in all cases.

All disks were observed at several wavelengths with similar angular resolution, from the optical to the millimeter range. This enables a comparison of the radial extent of different grain populations in the disks~(i.e., grain sizes). We assumed that 
the small dust responsible for the scattered light in the optical is tracing closely the gas distribution. Most disks have larger radial sizes in the optical-NIR than at millimeter wavelengths indicative of dust radial drift, the larger particles having drifted inward. Three of the disks have the same radial extent in both millimetric bands; these ones also have the sharpest apparent outer edges~(estimated between 20\% and 80\% of the peak flux): $\Delta r/r \sim 0.3$.  
Four sources have band~7 emission which is radially more extended than band~4, by about~12\% on average. 
However, current radial drift models predict larger differences - both between optical and millimeter, and between band~7 and band~4 - than we actually observe. This suggests that other mechanisms such as pressure traps are likely present in these disks to slow down or halt the radial drift.

The peculiar viewing angle of the disks presented in this survey allows us to obtain more direct information on their vertical structures. First of all, the direct comparison of the ALMA observations with scattered light data shows that these disks have larger vertical sizes in the optical-NIR than at millimeter wavelengths, indicative of the different optical depths and of vertical dust settling. To further estimate the vertical distribution of millimeter grains~(parametrized as a "scale height"), we compared the shape of ALMA observations with four radiative transfer toy models of different mass, which include or not vertical settling. 
We computed the models at high inclinations, with the same angular resolution as our data, and considered two different values for the scale height of millimeter grains. We find that at least three disks of our survey require that the millimeter dust "scale heights" is low, of order of a few au at r=100\,au: these disks are vertically thin at millimeter wavelengths. This is much thinner than the gas traced by the small dust, which has a typical scale height of 10\,au~\citep{Burrows_1996, Stapelfeldt_1998}.  

On a case by case basis, for \hk, a more detailed comparison of the ALMA images with a published scattered light model~(re-computed at millimeter wavelengths) also suggests differences in vertical extent between millimeter and micron-sized grains, as previously suggested by~\citet{Duchene_2003}.  Also, for \tauZero, the only edge-on disk well resolved in the two millimeter bands, we find that the band~7 emission is about~1.5~times more extended vertically than the band~4. Assuming that the measured vertical extent is directly proportionnal to the dust scale height, this ratio is expected for relatively large dust grains in the simple 1-D diffusion theory or numerical simulations including non-ideal MHD effects~\citep{Dubrulle_1995, Riols_2018}, further supporting the idea that strong vertical dust settling has taken place, leading to an increase in dust concentration in the disk midplane. 

Finally, we find evidence of a more flared structure in \IRAS, suggesting that the millimeter grains in this Class~I source are less settled. The millimeter dust in this disk may be in transition between the vertically unsettled structures seen in some Class~0 objects, and the flatter dust found in our Class~II disks. 

In forthcoming studies, we will present the CO gas distribution measured in these disks. We will also produce more detailed case by case models of selected targets to quantify the dust and gas density distribution profiles.

\begin{acknowledgements}
The authors thank the referee for the constructive comments which improved significantly the paper. The authors also thank Laura Perez and Eric Villard for useful discussions. 
MV, FM, MB, GvdP, CP acknowledge funding from ANR of France under contract number ANR-16-CE31-0013. This project was financially supported in 2019 and 2020 by the CNRS as part of its programme 80|PRIME. GD and ZT acknowledge support from NASA under grant 80NSSC18K0442 and NSF grant AST-1518332. KRS acknowledges funding from the Space Telescope Science Institute under GO grant 12514.   CP acknowledges funding from the Australian Research Council via FT170100040 and DP180104235. FL acknowledges the support of the Fondecyt program N$\circ$3170360.
This paper makes use of the following ALMA data: {\scriptsize ADS/JAO.ALMA\#2016.1.00460.S, ADS/JAO.ALMA\#2016.1.01505.S, ADS/JAO.ALMA\#2016.1.00771.S and ADS/JAO.ALMA\#2013.1.01175.S}. ALMA is a partnership of ESO~(representing its member states), NSF (USA) and NINS~(Japan), together with NRC (Canada), MOST and ASIAA (Taiwan), and KASI (Republic of Korea), in cooperation with the Republic of Chile. The Joint ALMA Observatory is operated by ESO, AUI/NRAO and NAOJ.
\end{acknowledgements}

\bibliographystyle{aa}
\bibliography{biblio}

\begin{appendix}

\section{Beam sizes}
\label{sec:beam_sizes}

We present the beam sizes obtained for the different observations in Table~\ref{tab:beams}. The last column describes the beam size of the maps used to compute the brightness temperatures and the spectral index maps. 
\begin{table*}[ht!]
    \centering
    \caption{Beam sizes of our observations and restored beam used to compute brightness temperatures and spectral index maps.}
    \begin{tabular}{ccccccccc}
        \hline \hline
         Sources &\multicolumn{2}{c}{ B7 } & \multicolumn{2}{c}{ B6} &\multicolumn{2}{c}{ B4 } &\multicolumn{2}{c}{ Restored B4 \& B7}\\
         &FWHM (\arcsec)&PA ($^\circ$)&FWHM (\arcsec)&PA ($^\circ$)&FWHM (\arcsec)&PA ($^\circ$)&FWHM (\arcsec)&PA ($^\circ$) \\
         \hline
         \tauZero&0.12$\times$0.08& -13&0.37$\times$0.18& 38&0.11$\times$0.04& 33&0.12$\times$0.08& -13\\
         \hh &0.14$\times$0.13& 2&0.26$\times$0.18& -2&0.15$\times$0.12& 39&0.15$\times$0.12& 38\\
         \IRAS&0.30$\times$0.24& -17&&&0.09$\times$0.04& 34&0.30$\times$0.24& -17\\
         \hk&0.11$\times$0.07& 3&&&0.12$\times$0.04& 42&0.12$\times$0.08& 42\\
         \hv &0.11$\times$0.08& 4&&& 0.12$\times$0.04& 41&0.12$\times$0.08& -4\\
         \JTroisZeroSept&0.11$\times$0.07& -8&&&0.11$\times$0.04& 35&0.11$\times$0.11& 35\\
         \JDeuxDeux &0.12$\times$0.07& -19&&&0.10$\times$0.04& 30&0.12$\times$0.08& -19\\
         \irasGerrit&0.48$\times$0.33& 8&&&0.11$\times$0.04& 36&\\
         \Oph &&&0.23$\times$0.13& -70&&\\
         \Junun&0.48$\times$0.28& -8&&&\\
         \Junsois&0.47$\times$0.28& -10&&&\\ 
         \hhq&0.48$\times$0.28& -5&&&\\
        \hline
    \end{tabular}
    \label{tab:beams}
\end{table*}
\section{Brightness temperature cuts}
\label{sec:appendix_T}

\begin{figure*}
\begin{center}
    \includegraphics[width =18cm]{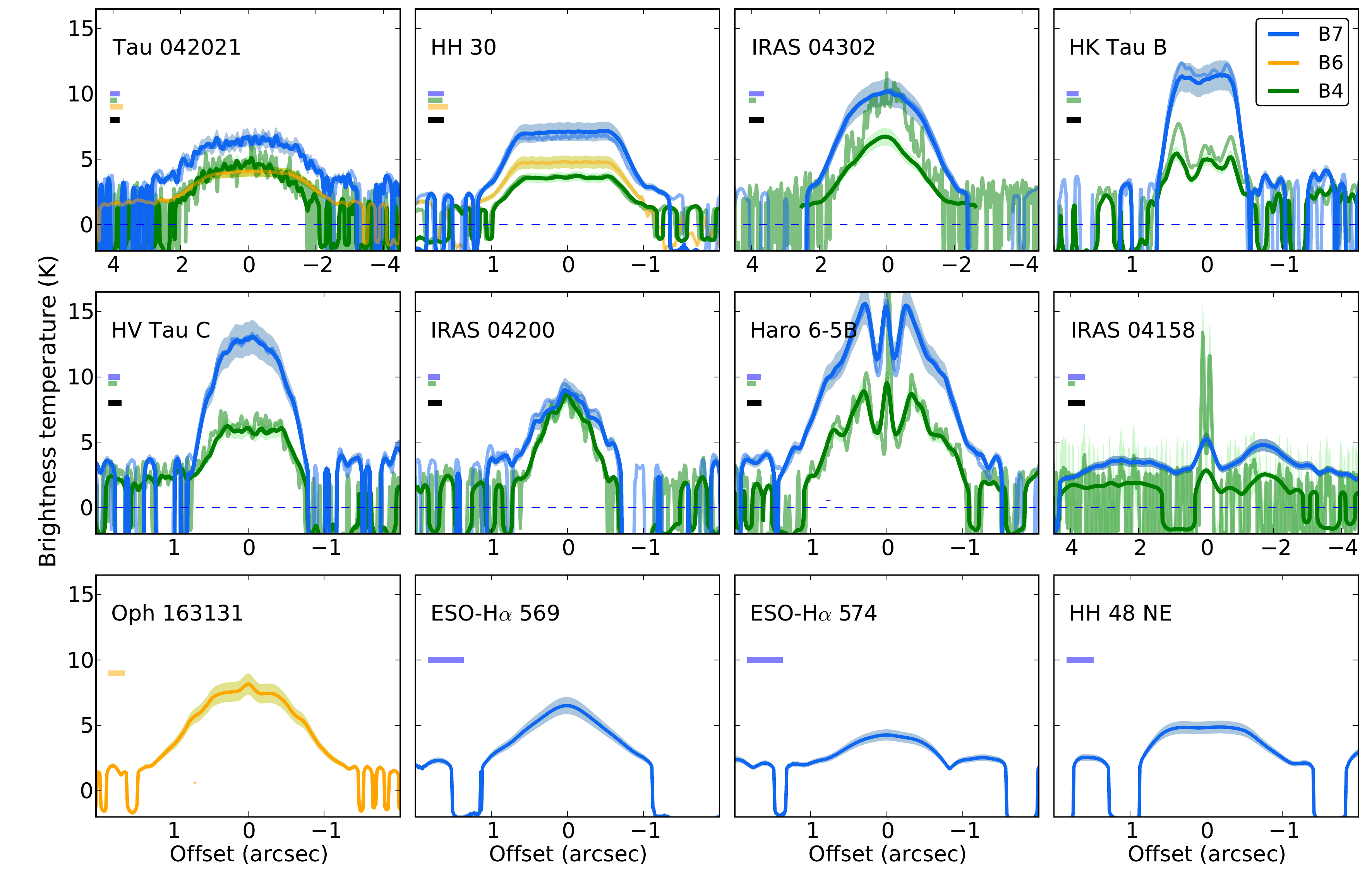}
    \caption{Observed brightness temperatures  as  a  function  of the  radial  distance  to  the  central  star. Solid lines profiles (band~7 and band~4 only) were computed with the same  angular resolution (restored beam in Table~\ref{tab:beams}), while the light lines have the original angular resolution. We report the beam sizes in the direction of the cut as horizontal lines in the left part of each plot. The restored beam corresponds to the black line.  Error bars correspond to 10\% of the brightness temperature.}
    \label{fig:brightness_T}
    \end{center}
\end{figure*}

 \begin{figure*}   
 \begin{center}
   \includegraphics[width =0.79\textwidth]{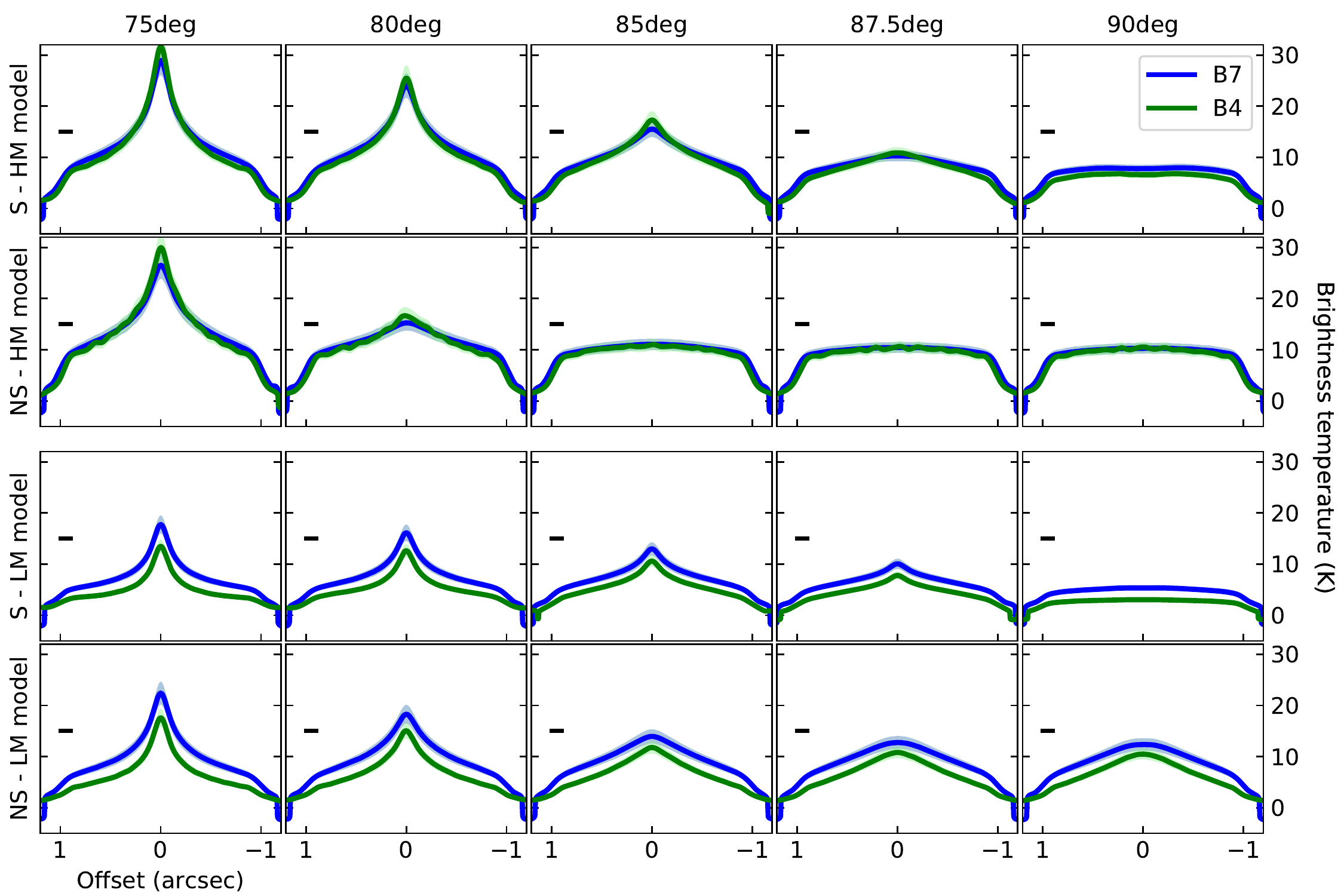}
    \caption{Brightness temperatures as  a  function  of the  radial  distance for the radiative transfer models presented in Section~\ref{sec:model}, computed at different inclinations. \emph{Top and third row:} High and low mass settling models, \emph{Second and bottom row:} High and low mass No Settling models. We report the beam size in the direction of the cut in the left part of each panel.}
    \label{fig:brightness_T_model}
    \end{center}
\end{figure*}

We present brightness temperature cuts along the major axis for all disks in Fig.~\ref{fig:brightness_T}. 
Solid lines represent the band~7 and band~4 cuts computed with the same beam, for which we reported the peak temperature in Table~\ref{tab:brightness_T}. 
We also show the brightness temperatures calculated with the resolution of the images presented in~Fig.~\ref{fig:alma_images} in light colors. We note that, as the beam sizes of the band~6 observations of \tauZero~and \hh~are significantly larger than the restored beam~(see Table~\ref{tab:beams}), it is not possible to compare directly the band~6 brightness temperatures to band~4 and band~7, the beam dilution will be different. Moreover, a direct comparison of brightness temperature between sources is difficult as they would have different levels of beam dilution. 

In Fig.~\ref{fig:brightness_T}, we find that most sources do not show strong differences in brightness temperatures between the different angular resolution studied. Only the band~4 observations of \JDeuxDeux~and \IRAS~show important variations of the brightness temperature with the beam size. The temperatures obtained with the restored beam are smaller than those with the original angular resolution. For both objects, this likely indicates that the disk~(or central point source for \JDeuxDeux) is significantly less resolved with the restored beam than at the original angular resolution. This leads to a lower brightness temperature due to beam dilution.\\

In Fig.~\ref{fig:brightness_T_model}, we present brightness temperature cuts at several inclinations computed for our radiative transfer models with and without including vertical settling. We also report the peak brightness temperatures obtained for our models in Table~\ref{tab:TbModel}. In each model, we find that as the inclination increases, the peak temperature decreases. This indicates that the disk becomes optically thicker with inclination. At the highest inclinations millimeter emission likely originates in the outer radii of the disk. We note that this is presumably also the case for several disks of our survey in which low brightness temperatures are measured even if they are well resolved~(for example for \tauZero).

Additionally, we find that, in the low mass models, the band~4 brightness temperatures are lower than the band~7 brightness temperatures. This is not the case for the high mass model where the band~4 brightness temperature can sometimes be higher than the band~7 temperature. 
In the observations, we detect large variations in brightness temperatures between bands~(Fig.~\ref{fig:brightness_T}, Table~\ref{tab:brightness_T}). 
While this suggest that the disks are optically thinner than the high mass models, further studies are needed to explain the differences in brightness temperature measured in the observations.

\begin{table*}[]
    \centering
    \caption{Peak brightness temperatures for our radiative transfer models}
    \begin{tabular}{ccccccccc}
        \hline \hline
         &\multicolumn{2}{c}{S-HM}  & \multicolumn{2}{c}{NS - HM}&\multicolumn{2}{c}{S-LM}  & \multicolumn{2}{c}{NS - LM} \\
         Incl ($^\circ$) &B7 (K)& B4 (K)& B7 (K)& B4 (K)&B7 (K)& B4 (K)& B7 (K)& B4 (K) \\
         \hline
         75 & 28.9 & 31.6 & 26.4 & 30.0& 17.7& 13.5 & 22.4 & 17.6 \\
         80 & 24.0 & 25.5 & 15.2 & 16.7 & 16.1 & 12.6 & 18.3 & 15.0\\
         85.0 & 15.5 & 17.3 & 11.0 & 11.1  & 13.0 & 10.6 & 13.9 & 11.8\\
         87.5 & 10.3 & 10.9 & 10.5 & 10.7 & 10.1 & ~7.8 & 12.7 & 10.8\\
         90.0 & ~8.0 & ~6.8 & 10.3 & 10.5  & ~5.4 & ~3.1 & 12.4 & 10.5\\
         \hline
    \end{tabular}
    \label{tab:TbModel}
\end{table*}

\section{Spectral index maps of the high mass settled model}
\label{apdx:models}
\begin{figure*}
    \centering
    \includegraphics[width =0.75\textwidth]{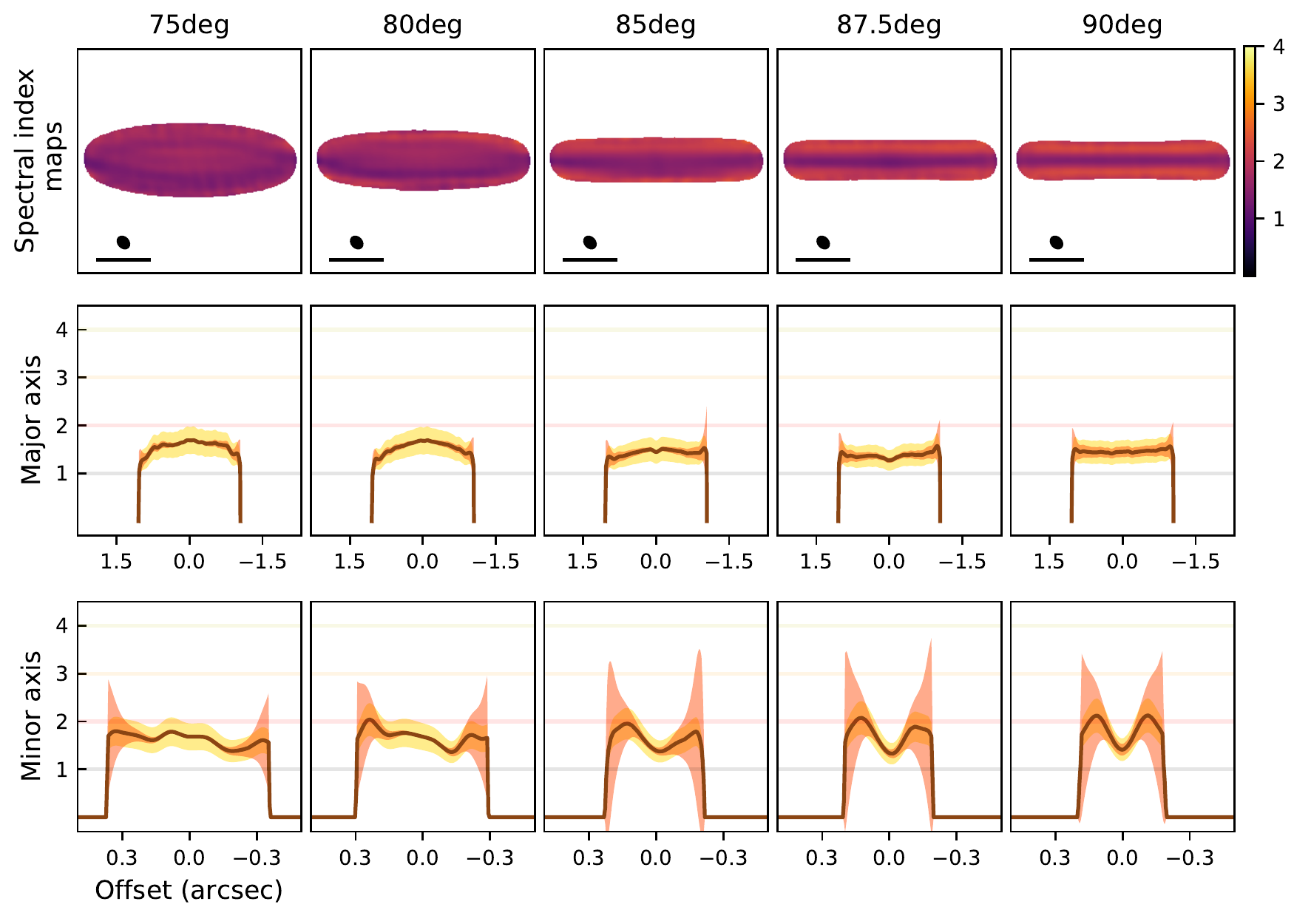}
    \caption{\emph{Top row:} Spectral index maps of the high mass settled model, applying a filter to keep only the pixels with more emission than 5$\sigma$ in both bands. The beam size is shown in the bottom left corner, along with a dark line representing a 0.5\arcsec\:scale. \emph{Middle row:} Spectral index cuts along the major axis. \emph{Bottom row:} Spectral index profiles along the minor axis, averaged along the major axis. For all cuts, yellow errors correspond to a flux calibration error of 10\% in both bands, while orange errors are estimated from the signal-to-noise in each band. The x-axis corresponds to the offset to the center of the disk in arcseconds.}
    \label{fig:alpha_model}
\end{figure*}

In order to interpret the spectral index maps shown in~Fig.~\ref{fig:alpha_map_cuts}, we computed the spectral index map of our high mass settled radiative transfer model presented in Section~\ref{sec:model}~(the most representative of our observations, see Section~\ref{sec:verticalExtent}), for inclinations between 75$^\circ$ and 90$^\circ$. We show the maps on the top row of~Fig.~\ref{fig:alpha_model}, and display the major axis cuts in the disk midplane in the middle row and the averaged minor axis profiles in the bottom row. As in Section~\ref{sec:vertical_extent}, we obtained the minor axis profiles over the full major axis for the disks that do not show ellipticity~(i.e., more inclined than 80$^\circ$), and over $\pm0.15$\arcsec\: for the least inclined models. 

The spectral index maps of the models show several features also seen in the observations. First, for most inclinations, it reaches values lower than two in the midplane~(as for \tauZero~and \JTroisZeroSept), which can indicate low temperatures or important scattering in the disk midplane. Second, at the highest inclinations, there is a slight increase in $\alpha_{mm}$ along the major axis, which is related to optical depth effects only. This variation is smaller than for several sources of our sample~(e.g., \tauZero, \IRAS, \JTroisZeroSept\:and \JDeuxDeux) which suggests that these disks are affected by radial drift. Finally, for inclinations larger than 85$^\circ$, the spectral index increases along the minor axis, similarly to what is measured in \tauZero, which is likely related to optical depth effect and vertical settling.

\section{Estimation of scattered light sizes}
\label{sec:estimation_hst_size}
\begin{figure}[]
\begin{center}
    \includegraphics[width =8cm]{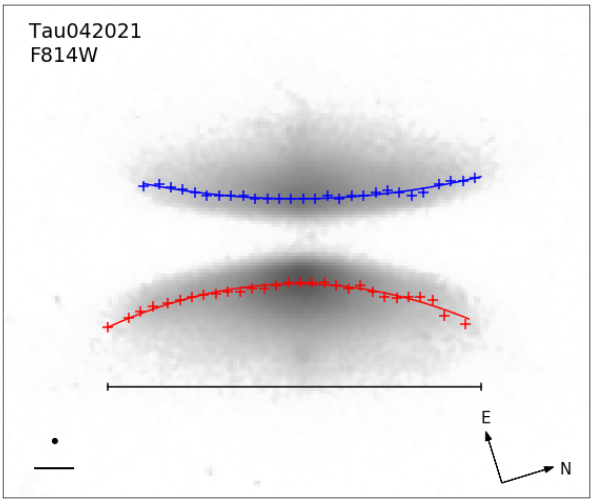}
    \caption{Example of scattered light size estimate in \tauZero. We represent the two spines of the nebula by the blue and red curves. The adopted radial size corresponds to the horizontal line, the maximum distance between spine points along the major axis.}
    \label{fig:hst_size}
    \end{center}
\end{figure}

Here we describe a method to estimate the radial extent of edge-on disks as observed in scattered light. 
To measure the extents of the scattered light images of the disks, we first find a set of points to define the spine of each nebula.
To this end, we extract cuts along the minor axis of the disk. When this cut contains two clear local maxima, we compute the centroid in a small region surrounding the peaks. If the two nebula are partially blended, or hard to disentangle, we perform a two-Gaussian fit, forcing the same FWHM of the two nebulae. The center of each Gaussian then defines the location of the spines. The process is initiated at the disk’s axis of symmetry and we proceed outward on each side until the surface brightness drops below $1-7\%$ of the peak surface brightness, depending on the signal-to-noise of the image and adjusted to match the visually detectable edge of the disk. The result is two sets of points that define the spines of the bright and counter nebulae~(blue and red points in Fig.~\ref{fig:alpha_model}). Outliers, due to bad pixels or substructure in the disk, are excluded based on deviations from a running median. We then fit second-order polynomial functions to each nebula, which provides adequate morphological information while minimizing sensitivity to noise and low-level departures from symmetry. The radial extent~(diameter) of the disk is then defined as the maximum distance between identified spine points along the major axis. We present an example of fit in Fig.~\ref{fig:hst_size}.

\end{appendix}
\end{document}